\documentclass[11pt]{article}
\usepackage{hyperref}
\usepackage{graphicx}
\usepackage{a4}
\usepackage{amsfonts}
\usepackage{pslatex}
\usepackage[latin1]{inputenc}
\usepackage[T1]{fontenc}

\newcount\printlevel
\def\rem#1#2{
\ifnum \printlevel>#1
#2
\fi
}
\def\log#1#2{
\ifnum \printlevel>3
{{\it #1:} #2 \\}
\fi
}%
\makeatletter
\def\@sect#1#2#3#4#5#6[#7]#8{\ifnum #2>\c@secnumdepth
  \def\@svsec{}\else 
  \refstepcounter{#1}\edef\@svsec{\csname the#1\endcsname.\hskip0.5em}\fi
  \@tempskipa #5\relax
  \ifdim \@tempskipa>\z@
    \begingroup 
      #6\relax
      \@hangfrom{\hskip #3\relax\@svsec}{\interlinepenalty \@M #8\par}%
    \endgroup
    \csname #1mark\endcsname{#7}\addcontentsline
      {toc}{#1}{\ifnum #2>\c@secnumdepth \else
        \protect\numberline{\csname the#1\endcsname}\fi #7}%
  \else
    \def\@svsechd{#6\hskip #3\@svsec #8\csname #1mark\endcsname
      {#7}\addcontentsline{toc}{#1}{\ifnum #2>\c@secnumdepth \else
        \protect\numberline{\csname the#1\endcsname}\fi #7}}%
  \fi \@xsect{#5}}
\@addtoreset{equation}{section}
\makeatother

\renewcommand\theequation{\ifnum \value{section}>0
 \arabic{section}.\arabic{equation}%
\else
\arabic{equation}%
\fi}

\def\cg{c_\Gamma}
\def\F#1#2{\,{{\vphantom{F}}_{#1}{F}_{#2}}}
\def\Li#1{\,{{\rm Li}_{#1}}}
\def\phpol{{\rm ph.\ pol.}}
\def\pol{\varepsilon}
\def\SpinSum{\widehat{\sum_{\st{pol.}}}}
\def\Tr{\mathop{\rm Tr}\nolimits}

\def\SUN{\ensuremath{\mbox{SU(N)}}}
\def\tr{\mbox{\sf Tr}}
\def\e{\epsilon}
\def\Se{{S_\e}}
\def\as{\alpha_s}
\def\gs{g_{\rm s}}
\def\Re{\mbox{Re}}
\def\smin{\ensuremath{s_{\st{min}}}}
\def\nn{\nonumber}
\def\st#1{\mbox{\scriptsize{\rm #1}}}
\def\dim{d}
\def\dlips{\ensuremath{dR^\dim}}
\def\dRcoll{dR^\dim_{\st{coll.}}}
\def\dRsoft{dR^\dim_{\st{soft}}}
\def\cA{{\cal A}}
\def\cC{{\cal C}}
\def\cS{{\cal S}}
\def\cN{{\cal N}}
\def\cO{{\cal O}}
\def\cV{{\cal V}}
\def\lc{{\mbox{\scriptsize lc}}}

\def\Dcg{\ensuremath{{\cal D}_{g\to g}}}
\def\Dcr{\ensuremath{{\cal D}^R_{g\to G}}}
\def\GB{\ensuremath{\Gamma^B(\phi \to g+X)}}
\def\GR{\ensuremath{\Gamma^R(\phi \to g+X)}}
\def\GRs{\ensuremath{\Gamma^R}}
\def\Gr{{\rm Gr}}
\def\Eq#1{eqn.~(\ref{#1})}
\def\msbar{\ensuremath{\overline{\mbox{MS}}}}
\def\L{\left(}\def\R{\right)}
\def\LB{\left[}\def\RB{\right]}
\def\tree{{\rm tree\vphantom{p}}}
\def\oneloop{{ \st{1-loop}}}

\def\Split{\mathop{\rm Split}\nolimits}
\def\soft{\mathop{\rm soft}\nolimits}
\def\Soft{\mathop{\rm Soft}\nolimits}

\def\Mixed{{\rm M}}

\def\Ctree{\Split^\tree}
\def\Ctreestar{\Split^{\tree\ast}}
\def\ConeR{\Split^{\oneloop\st{,R}}}
\def\Cone{\Split^\oneloop}

\def\Stree{\Soft^\tree}

\def\k#1{{k_{#1}}}
\def\la{{\lambda_a}}
\def\lb{{\lambda_b}}

\def\ll#1{{\lambda_#1}}
\def\h#1{{h^{(#1)}}}
\def\hR#1{{h^{R,(#1)}}}

\def\hr#1{{h_r^{(#1)}}}
\def\hv#1{{h_v^{(#1)}}}
\def\K#1{{K^{(#1)}}}

\def\Kr#1{{K_r^{(#1)}}}
\def\Kv#1{{K_v^{(#1)}}}
\def\P#1{{P^{(#1)}}}
\def\kj{{k_j}}
\def\kk{{k_k}}
\def\kl{{k_4}}
\def\tkl{{\tilde k_4}}
\def\tkijk{{\tilde k_{ijk}}}

\def\sijkl{s_{1234}}
\def\sijk{s_{123}}
\def\PP#1{\LB{#1}\RB_+}

\def\Color{\cite{BeGi87,KoLeNa88,MaPaXu88,BeKo91,DuDiMa00}}
\def\OneLoopFactorization{\cite{BeDiDuKo94,Ko99}}
\def\DoubleCollinear{\cite{CaGl98,CaGr98,Catani:1999ss,DuFrMa00}}

\begin{document}
\thispagestyle{empty}
\hfill{\parbox[t]{3.8cm}{%
    Saclay/SPhT--T03/085\\
    Karlsruhe/TTP03-16}}
\vspace*{2cm}
\begin{center}
  {\Large\bf
    Evolution Kernels from Splitting Amplitudes
    }
\end{center}
\vspace*{1cm}

\centerline{\large David A. Kosower${}^a$ and Peter Uwer${}^b$}
\vspace*{0.5cm}

\begin{center}
  \em
  $^a$Service de Physique Théorique${}^{\natural}$,
  Centre d'Etudes de Saclay,\\
  F-91191 Gif-sur-Yvette cedex, France\\
  $^b$Institut für Theoretische Teilchenphysik,
  Universität Karlsruhe\\ D-76128 Karlsruhe, Germany
\end{center}
\vspace*{0.5cm}

\begin{abstract}
We recalculate the next-to-leading order Altarelli--Parisi kernel using a 
method  which relates it to the splitting amplitudes describing the
collinear factorization properties of scattering amplitudes.  
The method breaks up the calculation of the kernel
into individual pieces which have an independent physical interpretation.
\end{abstract}

\vfill
${}^{\natural}$Laboratory of the
{\it Direction des Sciences de la Matière\/}
of the {\it Commissariat à l'Energie Atomique\/} of France.
\setcounter{page}{0}
\newpage
%
%
\section{Introduction}
\label{sec:Introduction}
The last quarter-century of experimental studies at colliders, 
complemented by theoretical investigations, have taught us 
that perturbative quantum chromodynamics (QCD)
gives an excellent description of the strong interaction probed at 
short distances.  
Indeed, the theory has reached a sufficient maturity that 
it is no longer the target of experimental studies,
but rather a tool in the search for new physics beyond the standard
model.

Upcoming collider experiments at the Tevatron and the LHC 
will require continuing refinements and progress in our 
ability to make precise predictions of QCD and
QCD-associated processes, along with an ability to give 
credible estimates of the associated uncertainties.  Recent
years have seen great progress in computing two-loop amplitudes
\cite{TwoloopFourPoint,TwoloopHelicity,TwoloopAllPlus}
in QCD, which is one of the ingredients essential to a next-to-next-to-leading
order (NNLO) description of collider processes.

The calculational framework for collider processes is based 
on our ability to compute short-distance matrix elements 
to increasing perturbative order in nonabelian gauge theories.  
It complements these process-dependent matrix elements with a
general understanding of factorization, which separates out 
the process-independent long-distance aspects.  
The long-distance aspects of scattering processes are
captured in the parton distribution functions of the 
scattering nucleon(s), and in fragmentation functions 
for identified outgoing hadrons.  Up to subleading power corrections, 
such functions along with the strong coupling $\alpha_s$ are
the only ingredients needed from outside perturbation theory for a description
of collider scattering processes.  

The parton distribution and fragmentation functions 
are functions of a momentum fraction, and of the scale $Q^2$ 
at which the hadron is probed.  As is well-known,
their values at different $Q^2$ are not independent, 
but are governed by the Altarelli--Parisi equation, 
whose kernel is computable in perturbation theory 
\cite{GrLi72a,GrLi72b,Li75,AlPa77,Do77,Pa80}.
It allows these functions to be evolved from a fixed scale $Q_0^2$ to other values
of $Q^2$.  The evolution kernels are known up to second order 
\cite{CuFuPe80,FuPe80,FlRoSa77,FlRoSa79,FlKoLa81a,FlKoLa81b,FlKoLa81c,Kalinowski:1981ju,Kalinowski:1981we,Gunion:1985pg}.
Curci, Furmanski, and Petronzio~\cite{CuFuPe80,FuPe80} used
the light-cone gauge, whereas Floratos et al.{}  obtained their 
results~\cite{FlRoSa77,FlRoSa79,FlKoLa81a,FlKoLa81b,FlKoLa81c}
  using a covariant gauge.
  Subtleties related to the proper renormalization of the gluon
  operator in covariant gauges were later clarified by Hamberg and
  van Neerven \cite{Hamberg:1992qt}. Based on this work the
second-order computation was rechecked more recently by Mertig 
and van Neerven,
who also computed the polarized kernels at this order
\cite{MevN96}. The calculation of the polarized kernels was 
checked by Vogelsang \cite{Vo96} using the light-cone gauge. 
The use of light-cone gauge was also investigated in
refs. \cite{Ellis:1996nn,Heinrich:1998kv,Bassetto:1998uv}.  Its use
beyond next-to-leading order is subtle~\cite{Leibbrandt} and not fully
tested. 
An NNLO calculation is being undertaken, using the operator approach, by
Moch, Vermaseren and Vogt \cite{Moch:2002sn}. Furthermore fits to the 
first moments of the evolution kernels
\cite{Larin:1997wd,vanNeerven:2000wp}
have already been used to determine the NNLO parton   distribution
functions \cite{Martin:2000gq,Alekhin:2001ih}. 
We believe it is useful to develop other methods
applicable to an NNLO calculation, and that is our purpose here.

Accordingly,
we wish here to introduce an alternative approach to computing the
kernel \cite{Kosower:1999wx}.
We will do so using gauge-independent quantities describing the collinear behavior of various
amplitudes.  This has the advantage of breaking down 
the computation into pieces
which themselves already have a meaningful interpretation.  
It also allows us to 
avoid complications
associated with prescriptions for light-cone gauge.  
A related analysis has been presented in refs.~\cite{deFlorian:2000pr,deFlorian:2001zd}.
In this paper, we focus our
attention on flavor-independent contributions to the time-like gluon kernel.

At present there exist two different methods to calculate the 
Altarelli--Parisi kernels.   To contrast previous methods with the
approach presented in the present work, let us consider the basic
ingredients of factorization.
The key feature of factorization in the QCD-improved parton model 
is the following: a hadronic cross section (more precisely, the leading-twist
contribution) can 
 be factorized into a hard scattering coefficient and 
parton distribution or fragmentation functions (so-called PDFs).
All information about the hard scattering process 
is contained in the hard-scattering coefficient
which is in the realm of short-distance physics and hence can be calculated entirely in
perturbative QCD.  Long-distance effects, including the matching of partonic
states to hadronic ones and vice versa, are captured by the PDFs.  These
functions can be expressed as expectation values of composite operators between
hadronic states.  The matrix elements of
the composite 
operators cannot be calculated in perturbative
QCD, although they are (in principle) amenable to non-perturbative techniques
such as lattice QCD.  On the other hand,
 these non-perturbative matrix elements are  
{\it universal\/}, that is process-independent, and hence can be measured in
one process and then used to obtain predictions for all other processes of
interest.  While the matrix elements themselves cannot be calculated
perturbatively, their scaling behavior {\it can\/} be, and it is this
scaling behavior which is captured by the Altarelli--Parisi equation.  Their
scaling behavior, or equivalently their anomalous dimensions, are determined
by the ultraviolet singularities of QCD corrections to the matrix elements.

The ultraviolet singularities, and hence the anomalous dimensions, are independent
of the choice of external state in the composite-operator matrix element.  For
calculational purposes, we can therefore replace a hadronic state with a partonic
one.  This yields one method (the `OPE approach') of calculating the evolution kernels.

The other method used in the literature is the so-called `infrared approach'.
Here one starts with unrenormalized quantities.  The calculation yields
singularities, in particular in the partonic, short-distance, cross section.
(The singularities are typically regulated using a dimensional regulator.)
One can distinguish between ultraviolet divergences and mass singularities,
related to the collinear emission from initial-state partons (or from
final-state partons fragmenting into identified hadrons). Soft
singularities cancel between virtual and real corrections. Ultraviolet
singularities are removed through the usual renormalization procedure,
that is are absorbed into the definition of the physical coupling in terms
of the `bare' coupling.  For mass singularities, the situation is more
complicated.  In the infrared approach, they are canceled by 
corresponding singularities in the `bare' PDFs.  Equivalently, they
are absorbed into the definition of the physical PDFs in terms of the
`bare' PDFs.  For this to be possible, the singularities must of course
possess a universal, process-independent form.  (They do.)

The mass singularities in the PDFs determine their evolution with respect
to changes in the reference scale, so the consistency of this approach
means that the singularities determine the evolution kernels.  In turn,
a determination of the remaining singularities in the ultraviolet-subtracted
hard-scattering coefficient thus also yields a calculation of the 
evolution kernels.  To determine these singularities, one could in
principle calculate the complete partonic cross section for a specific
process, and extract its singularities.  In general, this is a
formidable task.  
For example,
in order to derive just the leading-order
(LO) kernels one already needs to know the singularities of the next-to-leading
order (NLO) hard scattering coefficient. 

Fortunately, this task can be simplified.  
One way is to use the observation of
 Ellis, Georgi, Machacek, Politzer, and Ross
\cite{ElGeMaPoRo78, ElGeMaPoRo79}, that in axial gauge only a certain
class of diagrams contribute to the singularities. It is on this observation
that the derivation of Curci, Furmanski, and Petronzio 
\cite{CuFuPe80,FuPe80} is based.
Their derivation is closely connected to the use of
light-cone gauge whose treatment beyond next-to-leading order, as mentioned
earlier, is not fully tested.  

The method we shall use in the present paper is related but distinct.
We use the knowledge of factorization at the amplitude level,
more specifically the {\it splitting amplitudes}
which describe the collinear limits of scattering amplitudes. 
While it is intuitively clear that the factorization and universality
of the `mass' singularities in a hard-scattering cross section follow
from this factorization at the amplitude level, up to now no
explicit demonstration has been given in the literature.  We give
such a demonstration, using the phase-space slicing method, and
use the connection to derive the NLO evolution kernels.  
In the context of our derivation, the splitting
amplitudes represent smaller parts of the calculation that can
be verified independently.  Furthermore, since they describe properties
of on-shell amplitudes, they are manifestly gauge-invariant; the method
we describe is accordingly independent of the use of light-cone gauge.

Splitting amplitudes have been used extensively as a check on new
calculations of amplitudes, since the constraint of obeying the correct
behavior in all collinear limits is a strong one.  Indeed, the 
one-loop splitting
amplitudes have also been used, via such constraints,
 to give conjectured forms~\cite{AllPlus} for certain
classes of one-loop amplitude with an arbitrary number of external legs
(later proven by Mahlon~\cite{Mahlon}), and as an aid in the initial
derivation of
another all-$n$ class of amplitudes obtainable using the unitarity-based
method~\cite{BDDK}.

In this work we will focus on
 the  timelike $g\to gg$-kernel, with $g$ denoting a gluon. 
To simplify the derivation we will consider a gauge theory with no 
light fermions, and with an additional
massive colorless scalar coupled to the gluons via an
effective vertex. 

At leading order, the splitting amplitude can be regarded
as the amplitude for finding a parton of given momentum fraction 
inside a parent parton.  It is therefore natural to expect the
Altarelli--Parisi kernel, which describes the probability of 
finding a parton of given momentum fraction inside a parent parton,
to be the square of the splitting amplitude.  We make this correspondence
more precise in section~\ref{sec:LOsection}.  Beyond leading order,
we expect that there will be virtual corrections to this picture.
Indeed, they are given by the one-loop splitting amplitude, more
precisely by its interference with the leading-order splitting
amplitude.  As usual in gauge theories, however, there are additional
singularities due to the emission of soft or collinear partons, and
so we must also integrate over corresponding real-emission contributions.
Heuristically, these two contributions, less the iterated leading-order
kernel, give the next-to-leading order Altarelli--Parisi kernel.

The outline of the paper is as follows. In section
\ref{sec:Prerequisits} we review briefly
the factorization of color ordered amplitudes.
In section \ref{sec:Framework} we setup the general framework.
In the following section we illustrate 
the new approach by the rederivation of 
the leading-order kernel. We will discuss the leading-order derivation
in great detail because parts of it will be reused in the 
derivation of the NLO kernel which we present in section \ref{sec:NLOsection}.
We give our conclusions in section \ref{sec:Conclusion}.

\section{Collinear Factorization}
\label{sec:Prerequisits}
The properties of gauge theories are easiest to discuss
in the context of a color decomposition \Color.
At tree level, for all-gluon amplitudes
such a decomposition takes the form,
\begin{eqnarray}
  {\cal A}_n^{(0)}(\{k_i,\lambda_i,a_i\}) = 
  \sum_{\sigma \in S_n/Z_n} \Tr(T^{a_{\sigma(1)}}\cdots 
  T^{a_{\sigma(n)}})\,
  A_n^{(0)}(\sigma(1^{\lambda_1},\ldots,n^{\lambda_n}))\,,
\label{eq:TreeColorDecomposition}
\end{eqnarray}
where $S_n/Z_n$ is the group of non-cyclic permutations
on $n$ symbols, and $j^{\lambda_j}$ denotes the $j$-th momentum
and helicity.
We use the normalization 
\begin{equation}
  \Tr(T^a T^b) = \delta^{ab}
\end{equation}
for the generators
of \SUN. The same color decomposition as shown in 
\Eq{eq:TreeColorDecomposition} holds for the amplitudes we
shall consider, for the process $\phi\rightarrow g\cdots g$,
where $\phi$ denotes a colorless heavy scalar. 
One can write analogous formul\ae\ for amplitudes
with quark-antiquark pairs.
The color-ordered or partial amplitude $A_n$ is gauge invariant.
\begin{figure}[htbp]
  \begin{center}
    \centerline{%
      \includegraphics[width=0.7\textwidth,bb=60 336 521 456]{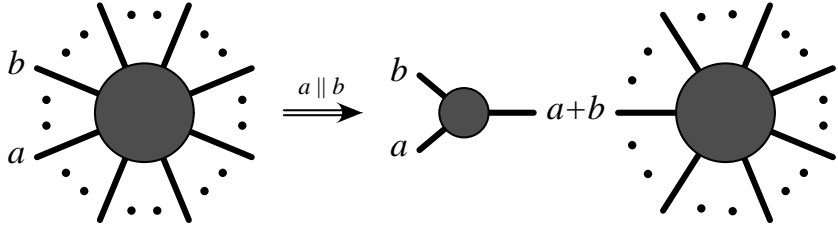}}
    \parbox{0.8\textwidth}{\caption{A schematic depiction 
        of the collinear factorization 
        of tree-level amplitudes, with the amplitudes labeled
        clockwise.}
      \label{fig:TreeFactorizationFigure}}
  \end{center}
\end{figure}
In the collinear limit, $k_a \parallel k_b$ of two adjacent legs,
the color-ordered amplitude $A_n$ is singular.  (It is finite
when the two collinear legs are not adjacent arguments, that is when
they are not color-connected.)  This
singular behavior has a universal form expressed by the 
tree-level factorization equation,
\begin{eqnarray}
  A_{n}^{(0)}(1,\ldots,a^\la,b^\lb,\ldots,n)  &&
  \stackrel{k_a\parallel k_b}{\longrightarrow}\nn\\
  &&\hspace{-2cm}g_s \sum_{\phpol\ \sigma}
 \Ctree_{-\sigma}(a^\la,b^\lb) 
    A_{n-1}^{(0)}(1,\ldots,(a+b)^\sigma,\ldots,n)+ \cdots,
\label{eq:CollinearA}  
\end{eqnarray}
where the dots represent terms which are finite in the limit.  In this equation, 
$\Ctree$ is the usual tree splitting amplitude, 
and the notation `$a+b$' means $k_a+k_b$. The QCD coupling is denoted
by $\gs$. The
notation `$\phpol$' indicates a sum over physical polarizations only.
(`Physical' here is in the sense of `transverse', and their number 
may depend
on the number of dimensions and on the variant of dimensional
regularization employed.) 
This factorization is
depicted schematically in fig.~\ref{fig:TreeFactorizationFigure}.
At tree level, one may derive the splitting amplitudes from
a string representation \cite{MaPa91} or from
the Berends--Giele recurrence relations \cite{BeGi88}.
It is characteristic of
gauge theories that the splitting amplitude has a square-root singularity,
$\Split \sim 1/\sqrt{s_{ab}}$, rather than a full inverse power of the
two-particle invariant $s_{ab}$.
Similar formul\ae\  hold in the triple-collinear case 
\DoubleCollinear.

\def\Gr{{\rm Gr}}
At one loop, the color decomposition 
analogous to~ \Eq{eq:TreeColorDecomposition} is
\begin{eqnarray}
  {\cal A}_n\L \{k_i,\lambda_i,a_i\}\R =
  \sum_J n_J
  \sum_{c=1}^{\lfloor{n/2}\rfloor+1}
  \sum_{\sigma \in S_n/S_{n;c}}
  \Gr_{n;c}\L \sigma \R\,A_{n;c}^{[J]}(\sigma),
  \label{eq:LoopColorDecomposition}
\end{eqnarray}
where ${\lfloor{x}\rfloor}$ is the largest integer less than or equal to $x$
and $n_J$ is the number of particles of spin $J$.
The leading color-structure factor,
\begin{equation}
  \Gr_{n;1}(1) = N\ \Tr\L T^{a_1}\cdots T^{a_n}\R\,,
\end{equation}
is just $N$ times the tree color factor, and the subleading color
structures are given by
\begin{eqnarray}
    \Gr_{n;c}(1) = \Tr\L T^{a_1}\cdots T^{a_{c-1}}\R\, \Tr\L
    T^{a_c}\cdots T^{a_n}\R.
    \label{eq:OneLoopColorStructures}
\end{eqnarray}
$S_n$ is the set of all permutations of $n$ objects,
and $S_{n;c}$ is the subset leaving $\Gr_{n;c}$ invariant.
The decomposition \Eq{eq:LoopColorDecomposition} holds separately
for different spins circulating around the loop.  The usual
normalization conventions take each massless spin-$J$ particle to have two
helicity states: gauge bosons, Weyl fermions, and complex scalars.
(For internal particles in the fundamental ($N+\overline{N}$) representation,
only the single-trace color structure ($c=1$) would be present,
and the corresponding color factor would be smaller by a factor of $N$.)

The subleading color amplitudes $A_{n;c>1}$ are in fact not independent
of the leading color amplitude $A_{n;1}\equiv A_n^{(1)}$.  
Rather, they can be expressed as
sums over permutations of the arguments of the latter \cite{BeDiDuKo94}.
(For amplitudes with external fermions, the basic objects are primitive
amplitudes~\cite{BeDiKo95} rather than the leading color one, 
but a similar dependence
of the subleading color amplitudes on the leading-color ones holds.)
\begin{figure}
\begin{center}
  \centerline{%
    \includegraphics[width=0.8\textwidth,bb=60 263 521 529]{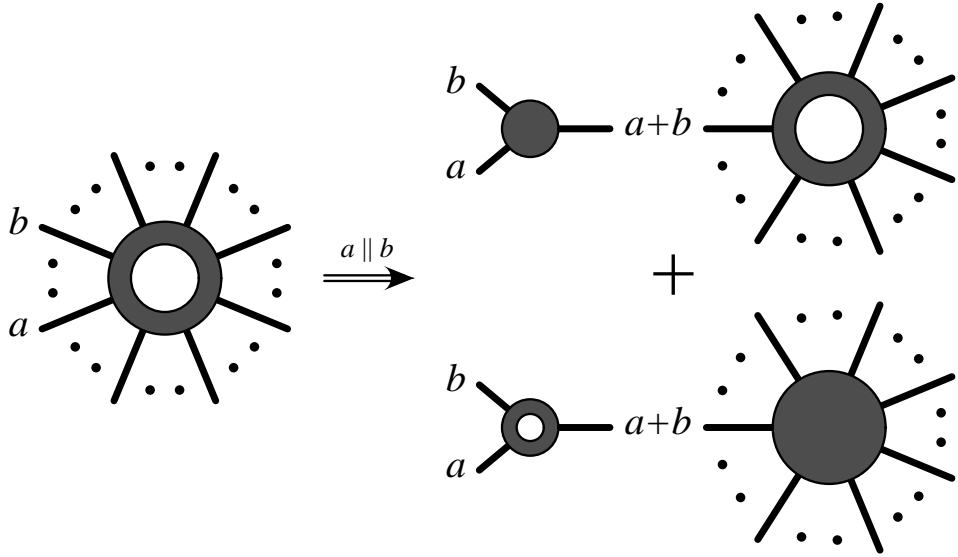}}
  \caption{A schematic depiction of the collinear factorization of one-loop
    amplitudes.   \label{fig:OneLoopFactorizationFigure}}
\end{center}
\end{figure}
As a result, it suffices to examine the collinear limits of leading color
amplitudes. The collinear limits of the subleading color then follow using this
relation.  The leading color one-loop amplitudes obey the following 
factorization~\OneLoopFactorization,
\begin{eqnarray}
  && A_n^{{(1)}}(1,\ldots,a^\la,b^\lb,\ldots,n) 
  \;{\buildrel a \parallel b\over{\relbar\mskip-1mu\joinrel\longrightarrow}}\cr
  &&\sum_{\phpol\ \sigma}  \biggl(
  \gs \Split^\tree_{-\sigma}(a^{\la},b^{\lb})\,
  A_{n-1}^{(1)}(1,\ldots,(a+b)^\sigma,\ldots,n)\nn\\
  &&\hskip 20mm  
  +\gs^3 \Split^\oneloop_{-\sigma}(a^\la,b^\lb)\,
  A_{n-1}^{(0)}(1,\ldots,(a+b)^\sigma,\ldots,n) \biggr) \;.
  \label{eq:OneLoopCollinearFactorization}
\end{eqnarray}
This factorization is
depicted schematically in fig.~\ref{fig:OneLoopFactorizationFigure}.

This form was originally deduced from explicit calculations of higher-point
amplitudes \cite{BeDiDuKo94,BeDiKo95}, 
but can also be proven more generally using the unitarity-based
method~\cite{Ko99}.  The latter proof also provides an explicit formula for the
one-loop splitting amplitude.  We used it 
\cite{KoUw99a} to calculate all the one-loop splitting amplitudes relevant in
QCD to all orders in the dimensional regularization parameter
\begin{equation}
  \e=(4-d)/2
\end{equation}
with $d$ being the dimensions of the spacetime.  A subset of
the terms higher order in $\e$ are needed for singular phase-space integrations
in NNLO jet calculations \cite{Weinzierl:2003ra}.
Bern et al. \cite{BernChalmers,BeDuSc98,BeDuKiSc99} derived
the one-loop splitting amplitudes 
from an analysis of one-loop integrals.

\section{Framework}
\label{sec:Framework}
In this section we derive a relation between the singularities in a
`partonic' (unsubtracted) cross section and the evolution
kernels.  We use this relation in following sections to 
compute the leading-order and the next-to-leading order kernels.
Our derivation follows closely the one given
in ref. \cite{CoSoSt}.   As mentioned earlier we restrict attention to
pure gluonic QCD together with an uncolored massive scalar $\phi$
 to which the gluons
couple via a higher-dimension operator. 
Such a coupling could, for example, be induced  via a 
heavy quark loop which is integrated out.   A similar derivation would of course
hold for a gauge theory with fermions as well.

Consider the production of a glueball 
$G$ in the decay of the massive scalar $\phi$. 
In order to calculate the corresponding decay rate or differential
distributions thereof, we need as ingredients 
the subtracted decay rate \GR\  for the 
production of gluons in the decay of the massive scalar, and the 
gluon-to-glueball fragmentation function \Dcr . 
Using these quantities the energy distribution of
the daughter glueball is given by the following expression,
\begin{equation}
  {d\Gamma(\phi\to G+X)\over dx_G}
  =
  {d\GR\over dx_g}
  \otimes 
  \Dcr
  \label{eq:Master}
\end{equation}
where the convolution `$\otimes$' is defined by
\begin{equation}
  [f\otimes g](x) = \int_0^1\int_0^1 dydz f(y) g(z) \delta(x-yz). 
\end{equation}
The energy fractions $x_G$ and $x_g$ are normalized to the 
mass $m_\phi$ of the scalar particle:
\begin{equation}
  x_{G,g} = {2 E_{G,g}\over m_\phi}, 
  \label{eq:energyfrac}
\end{equation}
where $E_G$ $(E_g)$ denotes the energy of the glueball (gluon).
We use the subscript $R$ to indicate that both the fragmentation 
function as well as the subtracted decay rate depend on the 
subtraction method used to define them; we have not put in an explicit
argument to show the
dependence on the factorization and the renormalization scale.
The universality of factorization allows us to write down
a formula for the `bare' decay rate \GB\ of the scalar
$\phi$ into an identified gluon in terms of the subtracted decay rate:
\begin{equation}
  {d\GB\over dx_g} =  {d\GR\over dx_g}
  \otimes 
  \Dcg.
  \label{eq:GluonicRate}
\end{equation}
The `bare' decay rate is unphysical, as it describes the `probability' of
finding a gluon with a given energy fraction inside a jet,
while of course a lone gluon is
not a colorless physical state.  
It will thus contain collinear or `mass' singularities, as will
$\Dcg$.   Our
purpose in considering $\GB$ is precisely to extract these singularities.
In the following we will assume that all singularities
(ultraviolet (UV), soft, and mass singularities)
are regulated  via a dimensional regulator.   The advantage of the bare
decay rate \GB\ is that it can be calculated purely in perturbative QCD.
As suggested by \Eq{eq:GluonicRate} the mass singularities which 
are present in \Dcg\  must exactly match those in the (unphysical)
partonic decay width \GB.  To see this, expand \Dcg\  in $\as$,
\begin{equation}
  \Dcg(z) = \delta(1-z) 
  + {\as\over 2\pi}\Dcg^{(1)}(z)
  + \L{\as\over 2\pi}\R^2 \Dcg^{(2)}(z)
  + O(\as^3),
\end{equation}
and then invert \Eq{eq:GluonicRate} to obtain,
\begin{eqnarray}
   {d\GR\over dx_g}
  &=&{d\GB\over dx_g}\otimes \Dcg^{-1}
  \nn\\
  &=&{d\GB\over dx_g}\otimes
  \bigg[\delta(1-z)-{\as\over 2\pi}\*\Dcg^{(1)}(z)\nn\\
  &&\hskip 5mm
  +\L {\as\over 2\pi} \R^2\*\L\Dcg^{(1)}\otimes\Dcg^{(1)}-\Dcg^{(2)}(z)
  \R 
  + O(\as^3)\bigg],
  \label{eq:FiniteGluonicRate}
\end{eqnarray}
where we have used the identity
\begin{eqnarray}
  \delta(1-z)&=&\LB 
  \delta(1-z) 
  + {\as\over 2\pi}\Dcg^{(1)}(z)
  + \L{\as\over 2\pi}\R^2 \Dcg^{(2)}(z)
  \RB\nn\\
  &\otimes&
  \LB\delta(1-z)-{\as\over 2\pi}\*\Dcg^{(1)}(z)
  +\L {\as\over 2\pi} \R^2\*(\Dcg^{(1)}\otimes\Dcg^{(1)}-\Dcg^{(2)}(z)
  ) 
  \RB   + O(\as^3)
\end{eqnarray}
to invert $\Dcg$.
The left hand side of \Eq{eq:FiniteGluonicRate} is a finite quantity,
and thus
the right hand side must be so as well.  If we now expand the `bare' partonic
decay width in $\as$, 
we see that order by order the singularities in the partonic 
decay width must be canceled by those which appear in $\Dcg^{(i)}$.
In particular, using
\begin{equation}
  {d\GB\over dx_g} = \h0(x_g) + {\as\over 2\pi} \h1(x_g) 
  + \L{\as\over 2\pi}\R^2 \h2(x_g)
  + O(\as^3),
\end{equation}
we obtain
\begin{eqnarray}
   {d\GR\over dx_g}
  &=& \h0+ {\as\over 2\pi} \LB\h1 - \h0 \otimes \Dcg^{(1)} \RB\nn\\
  &&\hspace{-2cm}+ \L{\as\over 2\pi}\R^2 \LB \h2 - \h1 \otimes \Dcg^{(1)} 
  + \h0\otimes \L \Dcg^{(1)}\otimes\Dcg^{(1)}-\Dcg^{(2)}(z)
  \R\RB+ O(\as^3).
  \label{eq:FiniteGluonicRateII}
\end{eqnarray}
(The reader may worry about powers of $\as$ implicit in the higher-dimension
operator coupling gluons to the heavy scalar.  Such factors are 
frozen at the scale where the operator is generated, and in any event do not
enter into the following arguments.)
The usual application of this expansion is to the calculation of the
differential decay width $d\GR/dx_g$: start with 
the `bare' partonic decay width $h^{(i)}$ and 
use \Eq{eq:FiniteGluonicRateII} (after ultraviolet subtractions as well) to obtain
the finite subtracted decay width, which predicts the decay of the scalar $\phi$ into
a glueball or more generally into hadrons. 
Here, we will use our knowledge of the collinear divergences in the
`bare' partonic decay width $\GB$, along with 
\Eq{eq:FiniteGluonicRateII}, to determine $\Dcg$.

Through ${\cal O}(\as^3)$, we can express $\Dcg$ in terms of the evolution 
kernels $\P0$ and $\P1$,
\begin{eqnarray}
  \Dcg &=& \delta(1-z)
  -{\as\over 2\pi} {1\over \e} \Se \P0(z)\nn\\
  &+&\L{\as\over 2\pi}\R ^2 \Se^2 \LB
  {1\over 2\e^2} \P0\otimes \P0
  +{1\over 4\e^2} \beta_0 \P0
  -{1\over 2}\* {1\over \e} \P1
  \RB + O(\as^3) .
  \label{eq:sven}
\end{eqnarray}
As usual $\beta_0$ denotes the first coefficient of the QCD 
$\beta$-function,
\begin{equation}
  \beta_0 = {1\over 3}\L 11N - 2 n_f\R
  \stackrel{n_f=0}{=}  {11  \over 3} N,
  \label{eq:Betafunction}
\end{equation}
and 
$\Se$ is the usual factor appearing in the modified minimal
subtraction (\msbar) scheme:
\begin{equation}
  \Se = (4\pi)^\e e^{-\e \gamma}
\end{equation}
with $\gamma$ the Euler constant. 
The general structure of this expansion follows 
from the renormalization-group
equation in the \msbar{} scheme, which we choose. 
In particular, using \Eq{eq:sven} we obtain for the $\mu$ dependence
of $\Dcr$: 
\begin{eqnarray}
   \mu^2{d\over d\mu^2} \Dcr
  &=& 
  - \Dcg \otimes \L \beta(\as,\e) {d\over d\as}   \Dcg^{-1}\R
  \otimes 
  \Dcr\nn\\
  &=& \LB {\as\over 2\pi}\P0 + \L{\as\over 2\pi}\R^2\P1 +\ldots\RB
  \otimes \Dcr
\end{eqnarray}
where we have used the $\beta$-function  in $\dim$
dimensions $\beta(\as,\e)$ in the \msbar{} scheme:
\begin{equation}
  \beta(\as,\e) = \mu^2{d\over d\mu^2} \as
  = -\as\*( \e + {\as\over 4\*\pi}\*\Se\*\beta_0) + \cO(\as^3).
\end{equation}
Inserting \Eq{eq:sven} in \Eq{eq:FiniteGluonicRateII} we obtain for
the subtracted differential decay rate,
\begin{eqnarray}
   {d\GR\over dx_g}
  &=& \h0 + {\as\over 2\pi} \LB\h1 
  + \h0 \otimes {1\over \e} \Se \P0(z) \RB
  + \L{\as\over 2\pi}\R^2 \bigg[ 
  \h2 + \h1 \otimes {1\over \e}\Se  \P0(z)
  \nn\\
  &&+\,\,\, \h0\otimes \L 
   {1\over 2\e^2} \Se^2 \P0\otimes \P0
  -{1\over 4\e^2} \Se^2 \beta_0 \P0
  +{1\over 2}\* {1\over \e} \Se^2 \P1
  \R\bigg].
  \label{eq:MSbarCounterterm}
\end{eqnarray}
The subtracted decay rate 
$\GR$ calculated via \Eq{eq:MSbarCounterterm} is the
\msbar-subtracted decay rate. In the following two sections we
illustrate how to calculate the divergences in $\h1$ and $\h2$, and
thereby determine the kernels $\P0$ and $\P1$.

\section{The Leading-Order Kernel}
\label{sec:LOsection}
In order to compute the leading-order kernel, $\P0$, we must isolate the
collinear singularities in $\h1$.  In terms of the matrix elements,
the leading-order partonic decay rate is given by the following formula,
\begin{eqnarray}
  \h0(x) 
  &=&
  \int \dlips(\k1,\k2)\; |{\cal A}^{(0)}(\phi\to g(\k1) g(\k2))|^2 
  [\delta(x_g - x_1) + \delta(x_g - x_2)]
\end{eqnarray}
where $\dlips(\k1,\ldots,k_n)$ denotes the phase-space measure 
in $\dim$ dimensions 
(including the symmetry factor $1/n!$) for $n$ gluons with momenta
$\k1,\ldots,k_n$,
\begin{equation}
  \dlips(\k1,\ldots,k_n) = {1\over n!} (2\pi)^\dim \delta(K-\sum_i k_i) 
  \prod_{i=1}^n {d^{\dim-1}k_i\over (2\pi)^{\dim-1} 2 k_i^0}.
\end{equation}
The leading-order amplitude for the 
decay $\phi\to gg$ is given
by ${\cal A}^{(0)}$. The energy fraction $x_i$ of the identified
gluon $i$ is defined
as in \Eq{eq:energyfrac} but with $E_{G,g}$ replaced by its energy $E_i$.
At next-to-leading order the partonic decay rate gets two additional
contributions, from virtual and real-emission corrections,
\def\mindent{\hskip 10mm}
\begin{eqnarray}
 {\as\over 2\pi} \h1(x_g)
  &=&
  {\as\over 2\pi} (\hv1(x_g)+\hr1(x_g)),\nn\\
  {\as\over 2\pi} \hv1(x_g)
  &=&  
  \int \dlips(\k1,\k2)\;
  2\Re[{\cal A}^{(1)}(\phi\to g(\k1) g(\k2)) 
  {\cal A}^{(0)\ast}(\phi\to g(\k1) g(\k2))]
  \nn\\ &&\mindent\times
  [\delta(x_g - x_1) + \delta(x_g - x_2)],\nn\\
  {\as\over 2\pi} \hr1(x_g)
  &=&  
   \int \dlips(\k1,\k2,\k3)\; |{\cal A}^{(0)}(\phi\to g(\k1) g(\k2) g(\k3))|^2 
  \nn\\ &&\mindent\times
  [\delta(x_g - x_1) + \delta(x_g - x_2)+ \delta(x_g - x_3)],
  \label{eq:NLO-cross-section}
\end{eqnarray}
where ${\cal A}^{(1)}(\phi\to gg)$ is the one-loop amplitude for the
decay $\phi\to gg$ while ${\cal A}^{(0)}(\phi\to g g g)$ is the tree-level
amplitude for the real-emission process $\phi \to ggg$. 

Both contributions in \Eq{eq:NLO-cross-section} 
contain collinear
singularities associated with the identified gluon
and will thus contribute to the kernel $\P0(z)$. 
(The individual contributions also contain soft and other collinear singularities.)
In the virtual corrections $\hv1$ the singularities arise from
the loop integration.   The general structure of infrared
divergences of one-loop amplitudes is known, 
we may extract them without further
knowledge of the specific process at hand.  This contribution will only
contribute to the term proportional to $\delta(1-z)$ in $\P0(z)$. 

Let us then turn to the contribution from real emission which 
will determine the `non-trivial' $z$ dependence of $\P0(z)$.  
(The $\delta(1-z)$ contribution can always be determined from the
$z$ dependent part through appeal to various sum rules, for example
that associated with momentum conservation.)
In the real emission contribution $\hr1$ the singularities
arise from the phase-space integration of singular terms in the 
matrix elements. These may be extracted by the 
phase space-slicing method \cite{GiGl92,GiGlKo93}. We will not need 
the whole apparatus developed for cross-section calculations, but only
the following basic elements.
The basic idea in the phase-space slicing method
is to split phase space into {\it resolved\/} and 
{\it unresolved\/} regions. In resolved regions, all outgoing partons
are `resolved', which is to say none become soft or collinear. 
The unresolved regions are the remaining regions of the phase space; in these
regions one or more final-state partons may be soft, or one or more pairs
may become collinear.

For the three-gluon final state, we may use the following partition of unity 
to separate the various regions,
 \begin{eqnarray}
  1 
  &=& 
  [\Theta(s_{12}-\smin) + \Theta(\smin-s_{12}) ]
  [\Theta(s_{23}-\smin) + \Theta(\smin-s_{23})]\nn\\
  &&[\Theta(s_{13}-\smin) + \Theta(\smin-s_{13})]\nn\\
  &=&
  \Theta(s_{12}-\smin)\Theta(s_{23}-\smin)\Theta(s_{13}-\smin)
  \quad\mbox{(1, 2, 3 hard)}\nn\\
  &+& \Theta(\smin-s_{12})\Theta(s_{23}-\smin)\Theta(s_{13}-\smin)
  \quad\mbox{(1, 2 coll.)}\nn\\
  &+&\Theta(s_{12}-\smin)\Theta(\smin-s_{23})\Theta(s_{13}-\smin)
  \quad\mbox{(2, 3 coll.)}\nn\\
  &+&\Theta(s_{12}-\smin)\Theta(s_{23}-\smin)\Theta(\smin-s_{13})
  \quad\mbox{(1, 3 coll.)}\nn\\
  &+& \Theta(\smin-s_{12})\Theta(s_{23}-\smin)\Theta(\smin-s_{13})
  \quad\mbox{(1 soft)}\nn\\
  &+&\Theta(\smin-s_{12})\Theta(\smin-s_{23})\Theta(s_{13}-\smin)
  \quad\mbox{(2 soft)}\nn\\
  &+&\Theta(s_{12}-\smin)\Theta(\smin-s_{23})\Theta(\smin-s_{13})
  \quad\mbox{(3 soft)}\nn\\ 
  &+& \Theta(\smin-s_{12})\Theta(\smin-s_{23})\Theta(\smin-s_{13})
  \quad\mbox{(`double unresolved')} ,
\label{eq:gggslicing}
\end{eqnarray}
with $s_{ij} = 2k_i\cdot k_j$.
The arbitrary parameter \smin\ introduced in \Eq{eq:gggslicing} 
controls the boundary between 
resolved and unresolved regions.  In parentheses we have classified 
the different contributions into resolved, soft and collinear contributions.
The `double unresolved' contribution does not contribute because 
it is kinematically forbidden. 
Thus only the contributions classified as soft or collinear  
will yield singularities. 

As we shall see, in fact
only the collinear contributions survive at the end after combining
virtual and real-emission contributions.  As one would expect, the 
singularities arising from soft regions cancel between these two types
of contributions. The evolution kernels are thus determined solely
by the collinear or `mass' singularities. From a practical point of view,
such cancellations are an important consistency check on the calculation. 

The cancellation of soft singularities can be seen heuristically
from the form of the partonic observable we are `measuring'.
For an $n$-parton final state, it takes the form,
\begin{displaymath}
  \cO_n = \sum_{i=1}^n \delta(x_g-x_i).
\end{displaymath}
We see that $\cO_n\to \cO_{n-1}$ when one gluon becomes soft, but this does
not happen when a pair becomes collinear.
  That is, the observable is soft-finite but not free of 
collinear singularities.  
Were the $\delta$-functions not present in our integral, the sum of the virtual and real
would of course be finite because it would just be the NLO correction to the
total decay rate.  With the $\delta$-functions present, but one of the gluons
soft, the corresponding term does not contribute, and all other terms are
insensitive to the soft momentum.  We can thus integrate the real-emission
contribution over the soft region, generating poles in $\e$ that cancels
the corresponding singularity in the virtual corrections.
In the case of collinear gluons, however, the two $\delta$-functions depending
on the momentum fractions of the collinear gluons will prevent
us from integrating over the appropriate region of phase space.  The
collinear divergence in the virtual corrections will therefore not be fully canceled.
The left-over singularity is precisely the term we are trying to extract.

We can use the symmetry of the matrix elements and of the phase-space measure
to restrict attention to the contribution where
gluons 1 and 2 are collinear.  The other two collinear contributions follow
through symmetrization.  Using the color decomposition introduced
in section \ref{sec:Prerequisits} we may write,
\begin{equation}
  {\cal A}^{(0)}(\phi\rightarrow g_1 g_2 g_3) = 
  \sum_{\sigma\in S_3/Z_3} 
  \tr(T^{a_{\sigma(1)}}T^{a_{\sigma(2)}}T^{a_{\sigma(3)}}) 
  A^{(0)}(\sigma(1),\sigma(2),\sigma(3)).
\end{equation}
Calculating the square of ${\cal A}^{(0)}(\phi\rightarrow g_1 g_2
g_3)$ yields different color structures.  For the determination of $\P0(z)$, it
is sufficient to consider the leading-color structure (denoted by the subscript `lc'):
\begin{eqnarray}
  \left.|{\cal A}^{(0)}(\phi\rightarrow g_1 g_2 g_3)|^2\right|_{\lc} 
  &=& 
  \left.\tr(T^a T^b T^c)\tr(T^c T^b T^a)
    \sum_{(ijk)= (123),(132)}
    |A^{(0)}(i,j,k)|^2\right|_{\lc}\nn\\
  &=&
  N^3
  (|A^{(0)}(1,2,3)|^2 + |A^{(0)}(1,3,2)|^2 ).
\end{eqnarray}
Within these terms, examine the contribution from the phase-space region where 
gluons~1 and~2 become collinear,
\begin{eqnarray}
   &&  \int \dlips(\k1,\k2,\k3)\;
   \left.|{\cal A}^{(0)}(\phi\rightarrow g_1 g_2 g_3)|^2\right|_{\lc}
  [\delta(x_g - x_1)+\delta(x_g - x_2)+\delta(x_g - x_3)] \nn\\
  &&\times\Theta(\smin-s_{12})\Theta(s_{23}-\smin)\Theta(s_{13}-\smin)\nn\\
  &=& 2\gs^2 N {1\over 3}\int\dlips(P,\k3) \dRcoll(\k1,\k2)\;
  | \cA^{(0)}(P,\k3)|^2 
   |\Ctree(\k1,\k2)|^2\nn\\
 && \hspace{-1.cm}\times[\delta(x_g - z x_P )+\delta(x_g - (1-z) x_P)+\delta(x_g - x_3)]
 \Theta(\smin-s_{12})\Theta(s_{23}-\smin)\Theta(s_{13}-\smin).
\label{eq:Collinear12}
\end{eqnarray}
The factor 2 in front accounts for the two different color orderings.
In deriving \Eq{eq:Collinear12} we have used the factorization
of the amplitudes as discussed in section \ref{sec:Prerequisits}. 
Using these results the factorization for the squared matrix element 
is given by
\begin{equation}
  |A^{(0)}(\k1,\ldots,k_i^{\lambda_i},k_j^{\lambda_j},\ldots,k_n)|^2
  \stackrel{k_i\parallel k_j}{\longrightarrow}
  \gs^2 |\Ctree_{-\lambda}(k_i^{\lambda_i},k_j^{\lambda_j})|^2
  |A^{(0)}(\k1,\ldots,(k_i+k_j)^\lambda,\ldots,k_n)|^2
\end{equation}
where we have eliminated cross terms of the form 
$\Ctree_{-\lambda}(k_i^{\lambda_i},k_j^{\lambda_j})
\times \L\Ctree_{\lambda}(k_i^{\lambda_i},k_j^{\lambda_j})\R^\ast $. 
These terms vanish upon azimuthal integration.
We have also made use of the factorization of the phase-space
measure in the collinear limit \cite{GiGl92},
\begin{eqnarray}
  \dlips(\k1,\k2,\k3) \stackrel{1\parallel 2}{\longrightarrow}
  {1\over 3} \dlips(P,\k3) \dRcoll(\k1,\k2)
  \label{eq:dRfac}
\end{eqnarray}
where
\def\collinearNorm{{\cal N}_c}
\begin{equation}
  \dRcoll(\k1,\k2) = \collinearNorm\,(\smin)^\e \,ds_{12}dz [s_{12} z(1-z)]^{-\e}
  \label{eq:CollinearPhasespace}
\end{equation}
and
\begin{equation}
  \collinearNorm={1\over16\*\pi^2}\*{1\over\Gamma(1-\e)}\*
  \L{4\*\pi\*\mu^2\over \smin}\R ^{\e}.
\end{equation}
We have defined $z$ to be the momentum fraction 
in the collinear limit,
\begin{equation}
  \k1 = z (\k1 + \k2) = z P,
  \quad\mbox{and}\quad
  \k2 = (1-z) (\k1 + \k2) = (1-z) P.
\end{equation}
We work throughout in $d=4-2\e$ dimensions.
The leading-order splitting amplitude is given by
\def\FirstTensor#1#2{%
-{\sqrt2\over s_{{#1}{#2}}} \L%
-\pol_{#1}\cdot\pol_{#2}\,k_{#2}\cdot\pol_P\,%
+ k_{#2}\cdot \pol_{#1}\, \pol_P\cdot\pol_{#2}\,%
- k_{#1}\cdot \pol_{#2}\,\pol_{#1}\cdot\pol_P\R}
\begin{eqnarray}
  \Ctree(1,2) = \FirstTensor12.
\end{eqnarray}
If we are only interested in the unpolarized kernels  we have to sum over the
final gluon polarizations and average over the fused-leg polarizations,
\begin{equation}
  \SpinSum |\Ctree(1,2)|^2
  \equiv
  {1\over 2} \sum_{\lambda,\ll1,\ll2} |\Ctree_{\lambda}(1^\ll1,2^\ll2)|^2
  ={2\over s_{12}}{(z^2-z+1)^2\over z\*(1-z)}
  \equiv  {2\over s_{12}} p(z).
  \label{eq:PolSumLO}
\end{equation}
The averaging depends in general on the variant of dimensional
regularization, this form holding for the conventional scheme \cite{CollinsCDR}.
Inserting \Eq{eq:CollinearPhasespace} and \Eq{eq:PolSumLO} 
in \Eq{eq:Collinear12}
we obtain
\begin{eqnarray}
  &&  -{4\collinearNorm\gs^2 N\over 3\e}\int \dlips(P,\k3)
  \int_{\tilde z(\k3,P)}^{1-\tilde z(\k3,P)} dz\; [z(1-z)]^{-\e}
  | \cA^{(0)}(P,\k3)|^2  p(z)\nn\\
  &&\mindent\times [\delta(x_g - z x_P)+\delta(x_g - (1-z) x_P)+\delta(x_g - x_3)]
  \Theta(s_{3P}-\smin),
  \label{eq:FirstLOForm}
\end{eqnarray}
with
\begin{equation}
  \tilde z(i,j) = {\smin\over s_{ij}}.
\end{equation}
\Eq{eq:FirstLOForm}
has almost the form we want: if we extend the
region of the $z$-integral by adding and subtracting 
\begin{eqnarray}
  &&-   {4 \collinearNorm \gs^2 N\over 3\e}\int \dlips(P,\k3) 
  \L \int_0^{\tilde z(\k3,P)}
  +\int_{1-\tilde z(\k3,P)}^1\R 
  dz\; [z(1-z)]^{-\e}
  | \cA^{(0)}(P,\k3)|^2  p(z)\nn\\
  &&\mindent\times[\delta(x_g - z x_P)+\delta(x_g - (1-z) x_P)]
  \Theta(s_{3P}-\smin),
  \label{eq:why-not-paul}
\end{eqnarray} 
we can write the expression in \Eq{eq:FirstLOForm} as a convolution plus 
an additional term: 
\begin{eqnarray}
  &&{1\over 3} {\as\over 2\pi } {1\over \e} [ \K0\otimes \h0](x_g) \nn\\
  &+&{2\over 3} N \int  dR_2^\dim(1,2)| \cA^{(0)}(1,2)|^2 
  [\delta(x_g - x_1)+\delta(x_g - x_2)]
    \cC^{(0)}(1,2,1),
    \label{eq:real-sing}
\end{eqnarray}
where
\begin{equation}
  \K0(z) =   16\*\pi^2\* \collinearNorm\* N\* \L 
  \delta(1-z) \cN
  - 2 \*(z\*(1-z))^{-\e}\* p(z)\R ,
\end{equation}
\begin{eqnarray}
  \cN &=& \int_{0}^{1}  dz z^{-\e} (1-z)^{-\e} p(z) 
  =
  -{3\*(1-\e)\*(4-3\*\e)
    \over 2\* \e\*(3-2\*\e)\*(1-2\*\e)}
  \*{\Gamma^2(1-\e)\over\Gamma(1-2\*\e)}\nn\\
  &=&
  -{2\over \e}-{11 \over 6} +({1 \over 3} \*\pi^2-{67 \over 18})\*\e
  + O(\e^2),
\end{eqnarray}
and
\begin{equation}
  \cC^{(0)}(i,j,k) = -2 {\collinearNorm\gs^2 \over \e}
  \int_{\tilde z(i,j)}^{1-\tilde z(k,j)} dz [z(1-z)]^{-\e}p(z).
\end{equation}
The other collinear regions -- in which $k_2\parallel k_3$ or $k_1\parallel k_3$ -- 
yield the same
result. In the sum of the three we get thus a factor of 3 which cancels
the factor $1/3$ from the phase space measure. 

Using the factorization of color-ordered amplitudes in the soft
limit \cite{MaPa91}
\begin{equation}
  A^{(0)}(1,\ldots,i^{\lambda_i},s^{\lambda_s},j^{\lambda_j},\ldots,n)
  \stackrel{s\st{ soft}}{\longrightarrow}
  \gs  N \Stree(i,s^{\lambda_s},j) 
  \times 
  A^{(0)}(1,\ldots,i^{\lambda_i},j^{\lambda_j},\ldots,n)
\end{equation}
together with the factorization of the phase space measure in the
soft limit \cite{GiGl92},
\begin{equation}
  \dlips(i,s,j) \stackrel{s\st{ soft}}{\longrightarrow}
   {1\over 3} \dRsoft(i,s,j) \dlips(i,j)
\end{equation}
with 
\begin{equation}
  \dRsoft(i,s,j) = \collinearNorm (\smin)^\e \L {s_{is} s_{js}\over s_{ij}}\R ^{-\e}
  {ds_{is} ds_{js}\over s_{ij}}
\end{equation}
we obtain (after relabeling)
\begin{eqnarray} 
  &&
  \int \dlips(\k1,\k2,\k3) 
  \left. |{\cal A}^{(0)}(\phi\to g(\k1) g(\k2) g(\k3))|^2\right|_{\st{lc}} 
  [\delta(x_g - x_1) + \delta(x_g - x_2)+ \delta(x_g - x_3)]\nn\\
  &&\times\Theta(\smin-s_{12})\Theta(s_{23}-\smin)\Theta(\smin-s_{13})
  +(\mbox{2 soft}) + (\mbox{3 soft}) \nn\\
  &=&
  \int \dlips(1,2) [\delta(x_g - x_1)+\delta(x_g - x_2)] 
  | \cA^{(0)}(1,2)|^2 2 N \cS^{(0)}(1,2)
\end{eqnarray}
for the singular contribution from the soft regions. The soft factor
$\cS^{(0)}(1,2)$ is given by
\begin{equation}
  \cS^{(0)}(1,2) = \sum_\lambda \int \dRsoft(2,3,1) 
  \gs^2|\Stree(2,3^\lambda,1)|^2 \Theta(\smin-s_{23})\Theta(\smin-s_{13}).
\end{equation}
As we shall see, we will not need the explicit result for $\Stree$.
Combining the soft and collinear contributions
we obtain 
\begin{eqnarray}
  \left. \hr1(x_g)\right|_{\st{lc, sing.}}
  &=&{\as\over 2\pi }{1\over \e}[ \K0\otimes \h0](x_g)\nn\\
  &&\hspace{-1.5cm}+
  \int \dlips(1,2) [\delta(x_g - x_1)+\delta(x_g - x_2)] 
  | \cA^{(0)}(1,2)|^2 2 N (\cS^{(0)}(1,2) + \cC^{(0)}(1,2,1))
  \label{eq:soft+collinear}
\end{eqnarray}
for the singular part of the real emission contribution.
To complete our analysis of the 
singular part in $\h1$ it remains only to add in the contribution from
the virtual corrections. 
The color decomposition of the one-loop amplitudes in pure gluonic QCD
is given by (c.f. section \ref{sec:Prerequisits}) 
\begin{equation}
  {\cal A}^{(1)} (\phi\rightarrow g_1 \cdots g_n) =
  \sum_{\sigma \in S_n/S_{n;c}}
     \Gr_{n;c}\L \sigma \R\,A_{n;c}(\sigma(1),\ldots, \sigma(n)),
\end{equation}
with $\Gr_{n;c}(1)$  defined in \Eq{eq:OneLoopColorStructures}.
Once again we are only interested in the 
leading color-structure. So it is sufficient to include 
$\Gr_{n;1}(1) A_{n;1}(1,\ldots,n)\equiv N\ \tr\L T^{a_1}\cdots
T^{a_n}\R A_{\lc}^{(1)}(1,\ldots,n)$ in our analysis.
The singularity structure of the one-loop color-ordered
amplitude $A_{\lc}^{(1)}(1,2)$ is known \cite{GiGl92,KunsztVirtual}.
Alternatively, we may reason as follows: 
suppose we are calculating the real-emission contribution to the
total $\phi$ decay rate. In this case we must replace the sum over 
$\delta$-functions by 1 in the derivation above.
We see immediately that the singular 
contribution of the soft and collinear regions to the total decay
rate reduces to 
\begin{equation}
   \int \dlips(1,2) 
  | \cA^{(0)}(1,2)|^2 2 N (\cS^{(0)}(1,2) + \cC^{(0)}(1,2,1)).
\end{equation}
On the other hand, unitarity \cite{LeeNauenberg,Kinoshita} dictates that
the total decay rate must be free of soft and collinear singularities,
order by order in perturbation theory.  (Note that the $\smin$-dependent
terms on the right-hand side are finite.)
This implies that the singular contribution from the leading-color
one-loop amplitude
satisfies
\begin{equation}
  N\left.2\Re(A_\lc^{(1)}(1,2) 
    A^{(0)\ast}(1,2))\right|_{\st{sing.}}
  =
  \left.
  -| A^{(0)}(1,2)|^2 2 N (\cS^{(0)}(1,2) + \cC^{(0)}(1,2,1))
   \right|_{\st{sing.}}.
\end{equation}
As a consequence we see that the singular contribution
from the virtual correction cancels the second term in \Eq{eq:real-sing}.
We thus obtain
\begin{equation}
  \left. \h1(x_g) \right|_{\st{lc, sing.}}
  = 
  {1\over \e} [ \h0 \otimes \K0 ](x_g)
\end{equation}
for the surviving singular term.
Note that $\h0$ should not be expanded in $\e$.
Comparing with \Eq{eq:MSbarCounterterm} we find
\begin{equation}
  \P0(z) = 
   -\left.\K0(z)\right|_{d=4} =   
  2 N \L {11 \over 12} \delta(1-z)
  + {1\over z} + \LB{1\over 1-z}\RB_+ - z^2 + z -2 \R ,
\label{eq:LeadingOrderResult}
\end{equation}
where we have used
\begin{equation}
  (1-z)^{-1-\e}=
  -{1\over \e}\delta(1-z)
  +\PP{{1\over 1-z}} 
  -\e\*\PP{{\ln(1-z) \over 1-z}}
+\, O(\e^2).
\end{equation}
In these equations, the plus prescription defines distributions via,
\begin{equation}
\PP{F(z)} = \lim_{\eta\rightarrow 0}
  \left\{ \Theta(1-z-\eta) F(z) - \delta(1-z-\eta) \int_0^{1-\eta} F(y) dy
   \right\},
\end{equation}
so that if $g(z)$ is well-behaved at $z=1$, then
\begin{eqnarray}
\int_x^1 dz\; {g(z)\over (1-z)_+} &=& \int_x^1 {g(z)-g(1)\over 1-z} + g(1) \ln(1-x),\\
\int_x^1 dz\; g(z)\Bigl[{\ln(1-z)\over 1-z}\Bigr]_+ &=& 
   \int_x^1 {(g(z)-g(1))\ln (1-z)\over 1-z} + {g(1)\over2} \ln^2(1-x).
 \label{PlusPrescriptionExample}
\end{eqnarray}

The phase-space slicing used here shares many features with that
used for the calculation of a typical infrared-finite observable.
In contrast to those observables, which necessarily allow the 
recombination of collinear partons in the real-emission terms,
and hence allow integrating over collinear phase space in a process-independent
way \cite{GiGl92}, the partonic decay rate \GB\ describes an
unphysical `probability' of finding a gluon with a given energy fraction 
inside the jet.  
This does not allow the
recombination of two collinear gluons to a hard one, 
and a further convolution with a physical
state distribution function is necessary to produce an infrared-finite observable.  The
uncanceled singularity does have a universal form, however, which is
why this `unphysical' object can be used to extract it.  
The singularities associated with soft-gluon emission do cancel,
because too soft a gluon won't contribute to the energy
fraction of any final-state hadron, and hence will drop out of
the calculation.

\section{Next-to-Leading Order Kernel}
\label{sec:NLOsection}
In order to calculate the NLO kernel we must study the singularities
in the NNLO decay rate. Again keeping only the contribution dominant 
in the number of colors (denoted by the subscript lc)
the decay rate is,
\begin{eqnarray}
  &&\left.{d\GB\over dx_g}\right|_{\st{NNLO, lc}}
  =\nn\\
  &&
  N^4\int \dlips(1,2)\;
  2\Re({A}_\lc^{(2)}(1,2) {A}^{(0)\ast}(1,2)
  )   \sum_{i=1}^2\delta(x_g - x_i) \nn\\
  &+&
  N^4 \int \dlips(1,2)\; |A_\lc^{(1)}(1,2)|^2 
  \sum_{i=1}^2 \delta(x_g - x_i) \nn\\
  &+&
   2 N^4 \int \dlips(1,2,3)\; 2\Re ({A}_\lc^{(1)}(1,2,3)
   {A}^{(0)\ast}(1,2,3))
  \sum_{i=1}^3 \delta(x_g - x_i) \nn\\
  &+&
   6 N^4 \int \dlips(1,2,3,4)\; |{A}^{(0)}(1,2,3,4)|^2
   \sum_{i=1}^4\delta(x_g - x_i).
  \label{eq:NNLO-cross-section}
\end{eqnarray}
The factors of $2$ and $6$ in front of the integrals in the last two terms
are combinatorial.  Once again we are interested only in the singular terms.
The singularities in the first two terms will contribute only
to the coefficient of $\delta(1-z)$ in the NLO kernel. 
The structure of the singularities,
needed for the direct computation of this coefficient
is now known
in a general two-loop 
amplitude~\cite{CataniConjecture,StermanTejeda,BDDSUSY}.
  However, the coefficient of the $\delta$-function
can always be computed using the sum rule constraints on the kernels.
We will therefore not compute it directly, and turn our attention
to the remaining terms in the kernel.  These are determined by the
three- and four-parton final states, which we discuss in the following
two subsections.

\subsection{Three-Parton Final State}
The treatment of this contribution is very similar to the treatment
of the three-parton final state in the computation of the leading-order kernel.
In particular, we may use the same phase-space slicing.
Here, we must consider the collinear limits of one-loop amplitudes in addition
to those of tree amplitudes.  Thus in addition to $\Ctree$, the one-loop splitting 
amplitude $\Cone$ also makes an appearance. 
Symmetry again allows us to focus on the configuration $k_1\parallel k_2$;
the other two collinear regions will give equal contributions.
 We are interested in the singular contribution arising from
\begin{eqnarray}
  &&
  2\int \dlips(1,2,3)\;\sum_{i=1}^{3}\delta(x_g - x_i)
  2 \Re(A_\lc^{(1)}(1,2,3)A^{(0)\ast}(1,2,3))\nn\\
  &&
  \mindent\times\Theta(\smin-s_{12})\Theta(s_{23}-\smin)\Theta(s_{13}-\smin).
\end{eqnarray}
Using the factorization of the phase-space measure
\Eq{eq:dRfac}, along with the factorization of the leading color 
one-loop amplitudes \OneLoopFactorization\  
(cf. \Eq{eq:OneLoopCollinearFactorization}),
\begin{eqnarray}
  A^{(1)}_\lc(1,\ldots,i^{\lambda_i},j^{\lambda_j},\ldots,n)
  &\stackrel{i\parallel j}{\longrightarrow}&
  \gs \Ctree_{-\lambda}(i^{\lambda_i},j^{\lambda_j})
  \times A_\lc^{(1)}(1,\ldots,(i+j)^\lambda,\ldots,n)\nn\\
  && +
  \gs^3 \Cone_{-\lambda}(i^{\lambda_i},j^{\lambda_j})
  \times A^{(0)}(1,\ldots,(i+j)^\lambda,\ldots,n),
\end{eqnarray}
we obtain
\begin{eqnarray}
  && 
  -{4\collinearNorm \gs^2\over 3\e} 
  \int \dlips(P,3) 
  \int_{\tilde z(3,P)}^{1-\tilde z(3,P)}dz\; [z(1-z)]^{-\e} p(z)
  [\delta(x_g - z z_P)+\delta(x_g - (1-z) z_P)]\nn\\
  &&\mindent\times 2   \Re[A_\lc^{(1)}(P,3)A^{(0)\ast}(P,3) ]\nn\\
  &&-{2\collinearNorm \gs^4\over 3\e}\,{\smin}^{-\e} \,
  \int \dlips(P,3) 
  \int_{\tilde z(3,P)}^{1-\tilde z(3,P)}dz\; [z(1-z)]^{-\e}
   [\delta(x_g - z z_P)+\delta(x_g - (1-z) z_P)]\nn\\
  &&\mindent\times 2 |A^{(0)}(P,3)|^2 
  \SpinSum\Re[(s_{12})^{1+\e} \Cone(1,2)\Ctreestar(1,2)]
  \nn\\  
  &&+{2\over 3}
  \int \dlips(P,3)\; \delta(x_g - x_3)\cC^{(0)}(3,P,3)
  2   \Re(A_\lc^{(1)}(P,3)A^{(0)}(P,3)^\ast )\nn\\
  &&+{2\over 3}
  \int \dlips(P,3)\;  \delta(x_g - x_3)\cC^{(1)}(3,P,3) |A^{(0)}(P,3)|^2 ,
\end{eqnarray}
where 
\begin{equation}
  \cC^{(1)}(i,j,k) 
  =
  -{\collinearNorm \gs^4\over 2\e}\,{\smin}^{-\e} \,
  \int_{\tilde z(3,P)}^{1-\tilde z(3,P)}dz [z(1-z)]^{-\e}
  2 
  \SpinSum\Re[(s_{12})^{1+\e} \Cone(1,2)\Ctreestar(1,2)].
\end{equation}
Note that 
\begin{equation}
   \SpinSum\Re[(s_{12})^{1+\e} \Cone(1,2)\Ctreestar(1,2)]
\end{equation}
does not depend on $s_{12}$. Extending the region of the $z$-integral
(by adding and subtracting the corresponding term), neglecting 
$\delta(1-z)$-type contributions, adding the contribution from
$k_2\parallel k_3$ and $k_1\parallel k_3$, and relabeling, we obtain
\begin{eqnarray}
 &&
 {\as\over 2\pi}
 N^4\int dz \int \dlips(1,2)\; [\delta(x_g - z x_1)+\delta(x_g - z x_2)]
 2   \Re[A^{(1)}(1,2)A^{(0)\ast}(1,2)]
 {1\over \e} \K0(z)
 \nn\\
 &&+
 \L {\as\over 2\pi} \R^2
 N^4\int dz \int \dlips(1,2)\; [\delta(x_g - z x_1)+\delta(x_g - z x_2)]
 |A^{(0)}(1,2)|^2
 {1\over \e} \Kv1(z)\nn\\
 &=& \L {\as\over 2\pi}\R^2
 {1\over \e}  [\hv1 \otimes \K0](x_g) 
 + \L {\as\over 2\pi}\R^2 {1\over \e} [\h0 \otimes \Kv1](x_g) 
 \label{eq:ThreePartonVirtualSingular}
\end{eqnarray}
where
\begin{equation}
  \Kv1(z)=
  -{1\over 2} (16 \pi^2)^2 \collinearNorm \, 
  \smin^{-\e}\,  N^2 (z(1-z))^{-\e}\SpinSum\Re((s_{12})^{1+\e} 
  \Cone(1,2)\Ctree(1,2)^\ast).
\end{equation}
The explicit results for $\Cone(1,2)$ 
can be found in refs. \cite{BeDiDuKo94,BeDuSc98,BeDuKiSc99,KoUw99a}:
\def\SecondTensor#1#2{{(k_{#1}-k_{#2})\cdot\pol_P\over\sqrt2 s_{{#1}{#2}}^2}%
\L s_{{#1}{#2}} \pol_{#1}\cdot\pol_{#2}%
- 2 k_{#2}\cdot \pol_{#1}\,k_{#1}\cdot \pol_{#2}\R}
\begin{equation}
  \Cone(1,2) = r_1(z) \Ctree(1,2) + r_2(z) \SecondTensor12
  \label{eq:ConeMaster}
\end{equation}
with
\begin{eqnarray}
  r_1(z) &=& {1\over 2} \L{\mu^2\over -s_{12}}\R^{\e}
  \LB z f_1(z) + (1-z) f_1(1-z)-2 f_2\RB,\nn\\
  r_2(z) &=&  {\e^2\over (1-2\e)(3-2\e)}
  \L{\mu^2\over -s_{12}}\R^{\e} f_2,
\end{eqnarray}
\begin{eqnarray}
  f_1(z) 
  &=& {2\over\e^2}\* \cg 
  \* \LB -{\Gamma(1-\e)\Gamma(1+\e)}
  \,z^{-1-\e} (1-z)^\e
  -{1\over z}
  +{(1-z)^\e\over z}\,\F21\L \e,\e;1+\e;z\R\RB\nn\\
  &=& 
  {2\over\e^2}\* \cg \* \LB -{\Gamma(1-\e)\*\Gamma(1+\e)} \* 
  \,z^{-1-\e} \* (1-z)^\e
  -{1\over z}
  +{(1-z)^\e\over z}
  +{\e^2\over z} \* \Li2(z) \RB + O(\e),\\
  f_2(z) &=& -{1\over\e^2}\* \cg,
  \label{eq:f1}
\end{eqnarray}
where $\F21$ is the Gauss hypergeometric function, 
$\Li2$ the dilogarithm, and the standard one-loop prefactor is,
\begin{equation}
\cg = {\Gamma(1+\e)\Gamma^2(1-\e)\over
  (4\pi)^{2-\e}\Gamma(1-2\e)}=\collinearNorm \L{\mu^2\over\smin}\R^{-\e} + O(\e^3).  
\end{equation}

Using the above results we find that
\begin{equation}
  r_1(z) = \collinearNorm \L{\smin\over -s_{12}}\R^{\e}
  \L 
  -{1\over \e^2}
  +{1\over \e}\*(\ln(z)+\ln(1-z) )
  -{1 \over 2} \*\ln(1-z)^2
  +\ln(z)\*\ln(1-z)-{1 \over 2} \*\ln(z)^2
  -{1 \over 6} \*\pi^2
  \R 
\end{equation}
and thus for the virtual contributions to $\K1$,
\begin{eqnarray}
  \Kv1(z) &=& 
  - (16\*\pi^2\* \collinearNorm)^2\* N^2\* \bigg\{
  p(z)
  \Bigl[-{1\over \e^2}
  +{2\over \e}\* (\ln(z)+\ln(1-z)\Bigr] \nn\\
  &&\mindent -2\*\ln(1-z)^2-2\*\ln(z)\*\ln(1-z)-2\*\ln(z)^2+{1 \over 3} \*\pi^2)
  +{1\over 6} 
  \bigg\}  + O(\e).
  \label{eq:Kv1}
\end{eqnarray}
The above equations give the unrenormalized splitting amplitude; the
renormalized one is, as given by \Eq{eq:ConeMaster} -- \Eq{eq:f1}
\begin{equation}
   \ConeR=\Cone 
  -{1\over 16\pi^2}
  \Se\*{1\over \e}\*{11\over 6}
  \Ctree,
  \label{eq:SplitRenormalized}
\end{equation}
which adjusts~\Eq{eq:Kv1} by,
\begin{eqnarray}
  \delta \Kv1(z) &=& 16 \pi^2 \collinearNorm \Se N^2 {11\over 3}
   \L{1\over \e}- \ln(z) - \ln(1-z)  \R p(z) + O(\e)\nn\\
   &=&
   (16\* \pi^2 \*\collinearNorm)^2  N^2 \L{\mu^2\over\smin}\R ^{-\e}  {11\over 3}
   \L{1\over \e}- \ln(z) - \ln(1-z)  \R \*p(z) + O(\e).
   \label{eq:Kv1ct}
\end{eqnarray}

\subsection{Four-Parton Final State}
We turn next to the four-gluon final state. 
The slicing of phase space is now more complicated than in the
three-gluon case.  A new feature arises: we can have {\it double unresolved}
contributions. Such contributions can originate from two soft gluons,
one soft gluon and a collinear pair, from two independent collinear pairs
or from a triple collinear configuration.
To derive a suitable slicing we start with the following partition of unity,
\begin{eqnarray}
 1
 &=&
 [\Theta(s_{12}-\smin) + \Theta(\smin-s_{12})]
 [\Theta(s_{23}-\smin) + \Theta(\smin-s_{23})]\nn\\
 &\times&[\Theta(s_{34}-\smin) + \Theta(\smin-s_{34})]
 [\Theta(s_{14}-\smin) + \Theta(\smin-s_{14})].
 \label{eq:pss-4gluons}
\end{eqnarray}
Each term in~\Eq{eq:NNLO-cross-section} is expressed in terms of a
single color-ordered amplitude, and hence 
the vanishing of an invariant $s_{ij}$ for non-adjacent gluons $i$, $j$
does not yield a singular contribution.
Expanding the right hand side of \Eq{eq:pss-4gluons}
we get a decomposition of the four gluon phase space into 
sixteen different regions.  We can classify these as chown in table 
\ref{tab:Classification}.
\begin{table}[!htbp]
  \begin{center}
    \leavevmode
    \renewcommand{\arraystretch}{1.3}
    \begin{tabular}[h]{cccc|l}
      \hline
      $s_{12}$ & $s_{23}$ & $s_{34}$ & $s_{41}$ &  {\it Comments\/}\\
      \hline
      $<$&$<$&$<$&$<$& \strut Triple-unresolved: kinematically forbidden.\\
      \hline
      $<$&$<$&$<$&$>$& \strut 1, 4 hard \& separated. $1\parallel 2$, 3 soft;
      or $3\parallel 4$, 2 soft;
      or 2, 3 soft.\\
      $<$&$<$&$>$&$<$& 3, 4 hard \& separated. $2\parallel 3$, 1 soft;
      or $1\parallel 4$, 2 soft;
      or 1, 2 soft.\\
      $<$&$>$&$<$&$<$& 2, 3 hard \& separated. $1\parallel 2$, 4 soft;
      or $3\parallel 4$, 1 soft;
      or 1, 4 soft.\\
      $>$&$<$&$<$&$<$& \strut 1, 2 hard \& separated. $2\parallel 3$, 4 soft;
      or $1\parallel 4$, 3 soft;
      or 3, 4 soft.\\
      \hline
      $<$&$<$&$>$&$>$& \strut 1, 3, 4 hard. $1\parallel 2 \parallel 3$; 
      or $1\parallel 3$, 2 soft; 
      or $1\not\parallel 3$, 2 soft.\\
      $<$&$>$&$>$&$<$& 2, 3, 4 hard. $2\parallel 3\parallel 4$; 
      or $2\parallel 4$, 3 soft;
      or $2\not\parallel 4$, 1 soft.\\
      $>$&$>$&$<$&$<$& 1, 2, 3 hard. $1\parallel 3\parallel 4$; 
      or $1\parallel 3$, 4 soft;
      or $1\not\parallel 3$, 4 soft.\\
      $>$&$<$&$<$&$>$& \strut 1, 2, 4 hard. $1\parallel 2\parallel 4$; 
      or $2\parallel 4$, 3 soft;
      or $2\not\parallel 4$, 3 soft.\\  
      \hline
      $<$&$>$&$<$&$>$& \strut All hard.  $1\parallel 2$, $3\parallel
      4$.  
      Double-collinear region.\\
      $>$&$<$&$>$&$<$& \strut All hard.  $2\parallel 3$, $1\parallel
      4$.  
      Double-collinear region.\\
      \hline
      $<$&$>$&$>$&$>$& \strut All hard.  $1\parallel 2$.  
      No contribution to NLO kernel.\\
      $>$&$<$&$>$&$>$& All hard.  $2\parallel 3$.  
      No contribution to NLO kernel.\\
      $>$&$>$&$<$&$>$& All hard.  $3\parallel 4$.  
      No contribution to NLO kernel.\\
      $>$&$>$&$>$&$<$& \strut All hard.  $1\parallel 4$.  
      No contribution to NLO kernel.\\
      \hline
      $>$&$>$&$>$&$>$& \strut All hard.  No singularities.  
      No contribution to NLO kernel.\\
      \hline
\end{tabular}
    \caption{Classification of the four parton phase space. The notation
      $<$ $(>)$ means that the corresponding invariant $s_{ij}$ is smaller
      (greater) than $\smin$.
      \label{tab:Classification}}
  \end{center}
\end{table}

In the first type of region, all invariants would be smaller
than $\smin$, but
this is kinematically forbidden, and so it gives rise to no contribution.
In the last type of region, all invariants are greater than $\smin$, so that
no singularities arise; we can set aside this region as well.  In the penultimate
type listed, one nearest-neighbor invariant is smaller than $\smin$ while the
other three are greater than it.  These give rise to single-unresolved
contributions for a collinear pair of gluons.  These regions give the NLO
corrections to processes with three distinct jets.  Their treatment follows
that of the single-unresolved regions in the three-parton final-state in
section~\ref{sec:LOsection}.  As given, they contribute only to the leading-order
kernel (and could be used to compute it had we not already used the NLO
decay rate to do so).  

It will nonetheless be convenient to add and subtract terms as done in
section~\ref{sec:LOsection} for the three-parton final state; the subtracted
terms will then enter into the calculation of the NLO kernel.  (This
amounts of course to shifting terms from other regions.)
The contribution from the $1\parallel 2$, for example, is given by
\begin{eqnarray}
  &&6 N^4  \int \dlips(1,2,3,4)\; |{A}^{(0)}(1,2,3,4)|^2
  \sum_{i=1}^4\delta(x_g - x_i)\nn\\
  &&  \mindent\times \Theta(\smin-s_{12})\*\Theta(s_{23}-\smin)
  \*\Theta(s_{34}-\smin)\*\Theta(s_{14}-\smin)\nn\\
  &=& 
  {N^3\over 2} \int \dlips(1,2,3)\;
  \int_{0}^{1} dz \,\,
  | A^{(0)}(1,2,3)|^2 {\as\over 2\pi} {1\over \e} \K0(z)\nn\\ 
  &&\mindent\mindent\times\sum_{i=1}^3 \delta(x_g - z z_i)
  \Theta(s_{12}-\smin)\Theta(s_{23}-\smin)\Theta(s_{13}-\smin)\nn\\ 
  &+& {3\over2} N^4 \int \dlips(\k1,\k2,\k3)\;
  \cC^{(0)}(3,1,2)
  | A^{(0)}(1,2,3)|^2  \nn\\
  &&\mindent\mindent\times [\delta(x_g -  x_1)+\delta(x_g - x_2)+\delta(x_g - x_3)]
  \nn\\ &&\mindent\mindent\times 
  \Theta(s_{21}-\smin)\Theta(s_{31}-\smin)\Theta(s_{23}-\smin),
\end{eqnarray}
where we have used the cyclic invariance of the color-ordered
amplitudes together with the freedom to relabel the momenta. 
We adjusted the boundaries of the $z$-integrals as 
described in section \ref{sec:LOsection}.
The last term will contribute only to
$\delta$-type terms; dropping it and including the contribution from 
$2\parallel 3$, $3\parallel 4$, $4\parallel 1$ we obtain,
\begin{eqnarray}
  && 2 N^3
  \int_{0}^{1} dz\int \dlips(1,2,3)\;
   \,\,
  | A^{(0)}(1,2,3)|^2 {\as\over 2\pi} {1\over \e} \K0(z)\nn\\ 
  &&\mindent\mindent\times\sum_{i=1}^3  \delta(x_g - z z_i)
  \Theta(s_{12}-\smin)\Theta(s_{23}-\smin)\Theta(s_{13}-\smin).
\end{eqnarray}
We can further rewrite it using~\Eq{eq:soft+collinear},
\begin{eqnarray}
  &&
  2N^3\int dz  \int \dlips(1,2,3)\; |{A}^{(0)}(1,2,3)|^2
  \sum_{i=1}^3\delta(x_g - x_i z) {\as\over 2\pi} {1\over \e} \K0(z) \nn\\ 
  &&
  \mindent\times\Theta(s_{12}-\smin)\Theta(s_{23}-\smin)
  \Theta(s_{13}-\smin)\nn\\
  &=&
  2N^3\int dz  \int \dlips(1,2,3)\; |{A}^{(0)}(1,2,3)|^2
  \sum_{i=1}^3\delta(x_g - x_i z) {\as\over 2\pi} {1\over \e} \K0(z)\nn\\
  && -\L{\as\over 2\pi}\R^2 {1\over \e^2}  [\h0 \otimes  \K0] \otimes \K0\nn\\
  && -
  N^2 \int \dlips(1,2)\; [\delta(y - x_1z)+\delta(y - x_2z)] 
  |A^{(0)}(1,2)|^2 
  2 (\cS^{(0)}(1,2) + \cC^{(0)}(1,2,1)) {\as\over 2\pi} {1\over \e} \K0(z)\nn\\
  &=&
   \L{\as\over 2\pi}\R^2 {1\over \e} \hr1\otimes\K0
  -\L{\as\over 2\pi}\R^2 {1\over \e^2} \h0 \otimes \K0 \otimes \K0 \nn\\
  &&-N^2
  \int \dlips(1,2)\; [\delta(y - x_1z)+\delta(y - x_2z)] 
  | A^{(0)}(1,2)|^2 
  2 (\cS^{(0)}(1,2) + \cC^{(0)}(1,2,1)) {\as\over 2\pi} {1\over \e} \K0(z).\nn\\
  \label{eq:OneCollinearPair}
\end{eqnarray}
The last term does not show a proper factorization because the 
combination $\cS^{(0)}(1,2) + \cC^{(0)}(1,2,1)$ is still process dependent,
but it will eventually cancel against virtual corrections with similar
structure.

We continue with contributions where two invariants are smaller than $\smin$.
There are two types, the third and fourth types in table \ref{tab:Classification}.  The latter,
\begin{eqnarray} 
  &&\Theta(\smin-s_{12})\*\Theta(s_{23}-\smin)
 \*\Theta(\smin-s_{34})\*\Theta(s_{14}-\smin)\nn\\
 &+&\Theta(s_{12}-\smin)\*\Theta(\smin-s_{23})
 \*\Theta(s_{34}-\smin)\*\Theta(\smin-s_{14}),
 \label{eq:TwoSmallDoubleCollinear}
\end{eqnarray}
correspond to configurations with two independent pairs of collinear gluons.
Here, the collinear integral over the variable not present in a $\delta$-function
can be done, while the other remains, so we obtain
\begin{eqnarray}
  \int dz\int dR_2^\dim(1,2)\cC^{(0)}(z k_1,k_2,(1-z) k_1)
  [\delta(x_g - x_1 z)+\delta(x_g - x_2 z)]|A^{(1)}(1,2)|^2
  {\as\over 2\pi}{1\over \e} \K0(z),
  \label{eq:TwoIndependentCollinearGluons}
\end{eqnarray}
where we have adjusted the boundaries of the $z$-integral as usual, and have
dropped terms which contribute only to the $\delta$-function in the kernel.
Note that the integral $\cC^{(0)}(z k_1,2,(1-z)k_1)$ still depends on the 
momentum fraction $z$,
\begin{equation}
  \cC^{(0)}(z k_1,k_2,(1-z)k_1) - \cC^{(0)}(k_1,k_2,k_1)
  = -   {\as\over 2\*\pi}\*{1\over\Gamma(1-\e)}\*
  \L{4\*\pi\*\mu^2\over \smin}\R ^{\e}  
  \L  {s_{12}\over \smin} \R^{\e}
  {1\over \e^2} \L z^\e + (1-z)^\e - 2 \R .
\end{equation}

In the third type of region in table \ref{tab:Classification}, two invariants with a
common momentum (`adjacent' invariants) are smaller than $\smin$,
\begin{eqnarray} 
 &&\Theta(\smin-s_{12})\*\Theta(s_{23}-\smin)
 \*\Theta(s_{34}-\smin)\*\Theta(\smin-s_{14})\nn\\
 &+&\Theta(\smin-s_{12})\*\Theta(\smin-s_{23})
 \*\Theta(s_{34}-\smin)\*\Theta(s_{14}-\smin)\nn\\
 &+&\Theta(s_{12}-\smin)\*\Theta(\smin-s_{23})
 \*\Theta(\smin-s_{34})\*\Theta(s_{14}-\smin)\nn\\
 &+&\Theta(s_{12}-\smin)\*\Theta(s_{23}-\smin)
 \*\Theta(\smin-s_{34})\*\Theta(\smin-s_{14})
 \label{eq:TwoSmallB}.
\end{eqnarray}
These contain subregions with qualitatively different types of
unresolved contributions, in which we must use different
factorizations -- as long as we are not using a limiting function 
unifying different limits (see for example ref. \cite{Kosower:2002su}).
For example, consider the region defined by
\begin{equation}
\Theta(\smin-s_{12})\*\Theta(\smin-s_{23})
 \*\Theta(s_{34}-\smin)\*\Theta(s_{14}-\smin).
 \label{eq:TwoSmallExample}
\end{equation}
Gluons $1$, $3$, $4$ cannot be soft, otherwise the last
two $\Theta$ functions would vanish. The constraint may be satisfied
in two distinct ways: gluon 2 can be soft, or three momenta can be
collinear, $1\parallel 2\parallel 3$ (with or without gluon being soft).

To distinguish the different subregions, we may introduce additional
$\Theta$ functions.  For example, multiply \Eq{eq:TwoSmallExample}
by 
\begin{equation}
  1 = \Theta(s_{123} -\smin) 
  +\Theta(\smin-s_{123})(\Theta(s_{24}-\smin)+\Theta(\smin-s_{24}))
\end{equation}
with $s_{ijk} = (k_i+k_j+k_k)^2$ to obtain
\begin{eqnarray}
   &&\Theta(\smin-s_{12})\*\Theta(\smin-s_{23})
 \*\Theta(s_{34}-\smin)\*\Theta(s_{14}-\smin)\Theta(s_{123} -\smin)
 \hphantom{+}\mbox{ 2 soft}\nn\\
 &&+\Theta(s_{34}-\smin)\*\Theta(s_{14}-\smin)\Theta(\smin-s_{24})
 \Theta(\smin-s_{123})
 \hphantom{\Theta(s_{14}-\smin)}\mbox{ $1\parallel 3$ and 2 soft}\nn\\ 
 &&+\Theta(s_{34}-\smin)\*\Theta(s_{14}-\smin)\Theta(s_{24}-\smin)
 \Theta(\smin-s_{123}).
 \hphantom{\Theta(s_{14}-\smin)}\mbox{ $1\parallel 2\parallel 3$}
 \label{eq:SlicingTwoSmall}
\end{eqnarray}
In the first term, gluons~1, 3, and~4 are resolved, while gluon~2
is soft, so this will contribute only to $\delta$-function terms.
In the second term, the matrix element will not be singular enough
to produce a pole unsuppressed by $\smin$ \cite{CaGl98}.
In the third term, we must use the triple-collinear factorization.

Before turning to the factorization and the computation of resulting
integrals, let us consider the last type of region (the second type in
table \ref{tab:Classification}), where three
nearest-neighbor invariants are smaller than $\smin$,
\begin{eqnarray}
  &&\Theta(\smin-s_{12})\*\Theta(\smin-s_{23})
 \*\Theta(\smin-s_{34})\*\Theta(s_{14}-\smin)\nn\\
  &+&\Theta(\smin-s_{12})\*\Theta(\smin-s_{23})
 \*\Theta(s_{34}-\smin)\*\Theta(\smin-s_{14})\nn\\
  &+&\Theta(\smin-s_{12})\*\Theta(s_{23}-\smin)
 \*\Theta(\smin-s_{34})\*\Theta(\smin-s_{14})\nn\\
  &+&\Theta(s_{12}-\smin)\*\Theta(\smin-s_{23})
 \*\Theta(\smin-s_{34})\*\Theta(\smin-s_{14} ).
\end{eqnarray}

To match the term we selected above, examine
\begin{equation}
  \label{eq:ThreeSmall}
  \Theta(s_{34}-\smin)\*\Theta(\smin-s_{14})
  \*\Theta(\smin-s_{12})\*\Theta(\smin-s_{23}).
\end{equation}
Once again we must introduce additional $\Theta$-functions to distinguish
between configurations in which different factorization formul\ae{} apply. 
In the case at hand, we must distinguish between
\begin{equation}
  1, 2 \mbox{ soft;} \quad
  1 \mbox{ soft},2\parallel 3;\quad  \mbox{or }
  2 \mbox{ soft}, 1\parallel 4.
\end{equation}
Multiplying \Eq{eq:ThreeSmall} by
\begin{equation}
  1
  =
  (\Theta(s_{13}-\smin) + \Theta(\smin-s_{13}))
  (\Theta(\smin-s_{24}) + \Theta(s_{24}-\smin))
\end{equation}
we obtain
\begin{eqnarray}
   && \Theta(s_{34}-\smin)\*\Theta(\smin-s_{23})
 \*\Theta(\smin-s_{12})\*\Theta(\smin-s_{14})\nn\\
 &&\mindent  \times
 \bigg(
   \Theta(\smin-s_{24})
 \Theta(\smin-s_{13}) \hphantom{)+\hskip 5mm}\mbox{ 1, 2 soft}\nn\\
 && \mindent\hphantom{\times\bigg()}+\Theta(s_{24}-\smin)
 \Theta(\smin-s_{13})
 \mbox{\hskip 5mm 1 soft, $2\parallel 3$}\nn\\
 &&\hphantom{\times\bigg()}\mindent+\Theta(s_{13}-\smin)
 \Theta(\smin-s_{24})
 \mbox{\hskip 5mm 2 soft, $1\parallel 4$}\bigg)
 \label{eq:SlicingThreeSmall1}
\end{eqnarray}
where we have dropped the contribution containing 
$\Theta(s_{13}-\smin)\Theta(s_{24}-\smin)$ because it is kinematically
forbidden.  The first term in \Eq{eq:SlicingThreeSmall1} contributes only
to $\delta$-function terms in the kernel, and we will not consider it further.

We will want to attach the first of the remaining two terms to the region
defined by \Eq{eq:SlicingTwoSmall}, while the last term we will associate
with a the similar product of theta functions with $(1,2,3,4)\rightarrow (4,1,2,3)$.
To organize this, define functions aggregating various theta functions
above,
\def\ThetaC{\Theta_C}\def\ThetaS{\Theta_S}\def\ThetaW{\Theta_W}
\begin{eqnarray}
\ThetaC(a,b,c;d) &=&
 \Theta(s_{ad}-\smin)\Theta(s_{bd}-\smin)\Theta(s_{cd}-\smin)\Theta(\smin-s_{abc}),\nn\\
\ThetaW(a,b,c) &=&  \Theta(\smin-s_{ab})\*\Theta(\smin-s_{bc})\Theta(\smin-s_{ac}),\nn\\
\ThetaS(a;b,c;d) &=& \Theta(\smin-s_{ad})\Theta(s_{bd}-\smin)\Theta(s_{cd}-\smin).
\end{eqnarray}

The complete surviving contribution of terms with two or three small invariants,
other than the double-collinear already accounted for 
in~\Eq{eq:TwoSmallDoubleCollinear}, is then given by inserting
\begin{eqnarray}
&&\sum_{\rho=(1234),(2341),(3412),(4123)} 
\hskip -5mm\bigl[\ThetaC(\rho_1,\rho_2,\rho_3;\rho_4) \nn\\
&&\hphantom{ \sum_{\rho=(1234),(2341),(3412),(4123)} }
  + \ThetaS(\rho_1;\rho_2,\rho_3;\rho_4)
                                       \ThetaW(\rho_1,\rho_2,\rho_3)  
\nn\\&&\hphantom{ \sum_{\rho=(1234),(2341),(3412),(4123)} }
   + \ThetaS(\rho_3;\rho_2,\rho_1;\rho_4)\ThetaW(\rho_1,\rho_2,\rho_3)
  \bigr]
\end{eqnarray}
into the integrand of the last term in~\Eq{eq:NNLO-cross-section}.

Each of the different permutations $\rho$
 corresponds roughly to a triple-collinear region,
with added regions where one of the `outer' gluons becomes soft.  
That is, they correspond to regions where a three-particle invariant
becomes small.  Let us focus
on one of these contributions, say as above the region where $s_{123}\rightarrow 0$.

Using the factorizations~\cite{CaGl98,CaGr98,Catani:1999ss}
\begin{equation}
  |A^{(0)}(1,2,3,4)|^2 \stackrel{1\parallel 2\parallel 3}{\longrightarrow}
  |\Ctree(1,2,3)|^2 \times |A^{(0)}(1+2+3,4)|^2
\end{equation}
and
\begin{equation}
   |A^{(0)}(1,2,3,4)|^2 \stackrel{1\soft, 2\parallel 3}{\longrightarrow} 
   \Mixed(1\soft, 2\parallel 3) \times |A^{(0)}(1+2+3,4)|^2
\end{equation}
the integrand in the region where $s_{123}\rightarrow 0$ is given by
\begin{eqnarray}
  I_{123} &=& |A^{(0)}(k_1+k_2+k_3,4)|^2 \nn\\
 && \times \bigg(\ThetaS(1;2,3;4)\ThetaW(1,2,3)\, \Mixed(1 \soft, 2\parallel 3)\nn\\
  && \hphantom{\times\bigg()} 
      +\ThetaS(3;2,1;4)\ThetaW(1,2,3) \, \Mixed(3 \soft, 1\parallel 2)\nn\\
  &&\hphantom{\times\bigg()} + \ThetaC(1,2,3;4)
 |\Ctree(1,2,3)|^2\bigg) + \mbox{ non-singular}.
\end{eqnarray}
As noted in ref. \cite{CaGl98} the function describing the mixed `soft-collinear' limit,
$\Mixed(a \soft, b\parallel c)$, 
can be derived from the triple-collinear splitting function.  Consequently,
the difference of the two limiting functions 
-- the soft-collinear one and the triple collinear one -- gives only
a finite contribution when integrated over a region where both are
valid.
In particular, as far as extracting poles is concerned, contributions
of the form 
\begin{equation}
  \ThetaS(1;2,3;4)
  \, (\Mixed(1 \soft, 2\parallel 3)-|\Ctree(1,2,3)|^2)
\label{eq:MixedTripleDifference}
\end{equation}
can be dropped.
We can use this to combine different contributions so as to simplify
the structure of the resulting integrals we must compute.  In general, 
the fewer constraints (theta functions), the better.

Using 
\begin{eqnarray}
 &&\Theta(s_{34}-\smin)\*\Theta(s_{14}-\smin)\Theta(s_{24}-\smin)
 =\nn\\
 &&1
 -\Theta(\smin-s_{14})\Theta(s_{24}-\smin)
 \Theta(s_{34}-\smin)
 -\Theta(s_{14}-\smin)\Theta(s_{24}-\smin)
 \Theta(\smin-s_{34})\nn\\
 &&-\Theta(\smin-s_{14})
  \Theta(s_{24}-\smin)\Theta(\smin-s_{34})
 -\Theta(\smin-s_{24})
\end{eqnarray}
along with~\Eq{eq:MixedTripleDifference},
we obtain
\begin{eqnarray}
  I_{123} &=& 
  \bigg(\Theta(\smin-s_{123}) |\Ctree(1,2,3)|^2\nn\\
  &&\hphantom{\bigg()}+\ThetaS(1;2,3;4)\*(\ThetaW(1,2,3)-\Theta(\smin-s_{123}))
  \, \Mixed(1 \soft, 2\parallel 3)\nn\\
  &&\hphantom{\bigg()}+\ThetaS(3;2,1;4)\*(\ThetaW(1,2,3)-\Theta(\smin-s_{123}))
  \, \Mixed(3 \soft, 1\parallel 2)\nn\\
  &&\hphantom{\bigg()}-\Theta(\smin-s_{14})
  \Theta(s_{24}-\smin)\Theta(\smin-s_{34})\Theta(\smin-s_{123})|\Ctree(1,2,3)|^2\nn\\
  &&\hphantom{\bigg()}-\Theta(\smin-s_{24})\Theta(\smin-s_{123})\|\Ctree(1,2,3)|^2
  \bigg) |A^{(0)}(1+2+3,4)|^2.
  \label{eq:ContributingTripleCollinearRegions}
\end{eqnarray}
The theta functions in the penultimate term require both gluons~1 and~3 to be soft,
and hence that term can contribute only to $\delta$-function terms in the kernel.
The theta function in the last term forces gluon~2 to be soft, and hence as discussed
above, the term will not contribute any singular terms.

One may also be tempted to drop the second and third terms.  This temptation should
be resisted, because although the region is small, the soft-collinear factorization
function is nonetheless sufficiently singular to produce a contribution, as shown
by a careful analysis in a different context~\cite{GeGl98}.

Let us first evaluate the primary contribution, given by the first term 
in~\Eq{eq:ContributingTripleCollinearRegions}.  The other regions related
by cyclic invariance (where respectively $s_{234}$, $s_{134}$, and $s_{124}$
vanish) give equal contributions; adding all four, we obtain
\begin{equation}
   24 N^4 \int \dlips(1,2,3,4)\; |{A}^{(0)}((1+2+3),4)|^2
  \sum_{i=1}^3 \delta(x_g - x_i) |\Ctree(1,2,3)|^2\Theta(\smin-s_{123}).  
\label{eq:TripleCollinearContribution}
\end{equation}
Factorizing  the phase-space measure in the triple collinear region 
\cite{GeGl98},
\begin{equation}
  \dlips(1,2,3,4) \stackrel{1\parallel 2\parallel 3}{\longrightarrow}
  {1\over 12}\dlips(P,4) \dRcoll(1,2,3),
\end{equation}
with
\begin{eqnarray}
  \dRcoll(1,2,3) 
  &=&
  {1\over 2^{8} \pi^{5}}
  {1\over \Gamma(1-2\e)}
  (4\pi)^{2\e}(-\Delta)^{-{1\over 2}-\e}
  dz_1dz_2dz_3  ds_{123}ds_{12}ds_{13}ds_{23}
  \nn\\
  && \delta(1-z_1-z_2-z_3)
   \delta(s_{123}-s_{12}-s_{13}-s_{23}),
   \label{eq:TripleCollinearPhasespace}
\end{eqnarray}
\begin{equation}
  \Delta 
  = 
  (z_3\*s_{12}-z_1\*s_{23} -z_2\*s_{13} )^2
  -4\*z_1\*z_2\*s_{23}\*s_{13},
\end{equation}
and
\begin{equation}
 k_i = z_i (\k1+\k2+\k3) = z_i P 
\end{equation}
the contribution of~\Eq{eq:TripleCollinearContribution} takes the form,
\begin{eqnarray}
  &&  2 N^4
  \int  \dlips(P,4) \dRcoll(1,2,3)\;
  |{A}^{(0)}((1+2+3),4)|^2
  \nn\\&&
  \hspace*{1cm}\sum_{i=1}^3 \delta(x_g - x_i) |\Ctree(1,2,3)|^2\Theta(\smin-s_{123})\nn\\
  &=& 
  N^2\int dz  \int  \dlips(1,2)  |{A}^{(0)}(1,2)|^2
  (\delta(x_g - z_1 z)+ \delta(x_g - z_2 z)) 
  \L{\as\over 2\pi}\R ^2 {1\over \e }\Kr1(z)\nn\\
  &=& \L{\as\over 2\pi}\R ^2 {1\over \e } [\h0\otimes\Kr1](x_g)
  \label{eq:TripleCollinear}
\end{eqnarray}
where
\begin{equation}
  \L{\as\over 2\pi}\R ^2 {1\over \e } \Kr1(z) = 
  N^2 \int \dRcoll(1,2,3) \sum_{i=1}^3 \delta(z-z_i)
  |\Ctree(1,2,3)|^2\Theta(\smin-s_{123}). \label{eq:K2real}
\end{equation}
Note that the factorization of the phase space given in 
\Eq{eq:TripleCollinearPhasespace} is exact up to an additional factor
$(1-{s_{123}\over s_{1234}})^{\dim-3}$
if one defines the `momentum fractions' outside the collinear region
by 
\begin{eqnarray}
  z_i =  {2 k_i\cdot\tkl\over 2\tkijk\cdot\tkl} 
  = {2 k_i\cdot\tkl\over \sijkl},\quad s_{ijkl} = (k_i+k_j+k_k+k_l)^2
\end{eqnarray}
with  $y=\sijk/\sijkl$ and 
\begin{eqnarray}
  \tkl = {1\over 1- y} \kl,\quad \tkijk = k_i + \kj + \kk - y \tkl.
\end{eqnarray}
Using \Eq{eq:TripleCollinearPhasespace} together with \cite{CaGl98},
\begin{eqnarray}
  \lefteqn{\sum_{\phpol}|\Ctree(1,2,3)|^2 = 2\times\bigg\{ }\nn \\
  && {(1-\e) \over  s_{12}^2 \* s_{123}^2} 
  \* {(z_2\*s_{123} - (1-z_3)\*s_{23})^2\over (1-z_3)^2}
  + {2\*(1-\e)\*s_{23}\over s_{12}\*s_{123}^2}
  + {3\*(1-\e)\over 2\*s_{123}^2}\nn\\
  &+& {1\over s_{12}\*s_{123}}
  \*\bigg( {(1-z_3\*(1-z_3))^2\over z_3\*z_1\*(1-z_1)} 
  - 2\*{z_2^2+z_2\*z_3+z_3^2\over 1-z_3}
        +{z_2\*z_1-z_2^2\*z_3-2\over z_3\*(1-z_3)} 
        + 2\* \e\*{z_2\over 1-z_3} \bigg)\nn\\
  &+& {1\over 2\*s_{12}\*s_{23}}
  \*\bigg(3\*z_2^2-2\*{(2-z_1+z_1^2)\*(z_2^2+z_1\*(1-z_1))\over z_3\*(1-z_3)} 
  + {1\over z_3\*z_1} 
  + {1\over (1-z_3)\*(1-z_1)}\bigg)
  \bigg\}\nn\\
  &+& (s_{12}\leftrightarrow s_{23}, z_1\leftrightarrow z_3),
  \label{eq:G-to-ggg}
\end{eqnarray}
(there is a factor of $1/4$ included here compared to
ref. \cite{CaGl98} to account for our 
normalization of the color matrices) we obtain 
\begin{eqnarray}
  \Kr1(z) &=&
  N^2 {1\over \Gamma(1-2\*\e)}\* \L {4\*\pi\*\mu^2\over \smin} 
  \* \R ^{2\*\e}\*\bigg\{
  -{5\over \e^2}\*p(z)
  +{1\over \e}\*\bigg(
  10\*p(z)\*\ln(1-z)
  -4\*(1+z-p(z) )\*\ln(z)\nn\\
  && -{1\over 6}\*{(-102\*z^3+55\*z^4+105\*z^2-102\*z+55)
    \over z\*(1-z) }\bigg)
  -{1\over 2}\* (p(z)-12\*(1+z))\*\ln(z)^2\nn\\
  &&-p(-z)\*S_2(z)
  -8\*(1+z)\*\Li2(z)
  -8\*p(z)\*\ln(z)\*\ln(1-z)\nn\\
  &&
  -10\*p(z)\*\ln(1-z)^2
  +{1\over 6}\*{73\*z^2+z+88\over z}\*\ln(z)\nn\\
  &&
  +{1\over 3}\*{-102\*z^3+55\*z^4+105\*z^2-102\*z+55\over z\*(1-z)}\*\ln(1-z)
  -{67\over 9}\*p(z) \nn\\
  && 
  -{1\over 36}\*{-330\*z+95+101\*z^2\over (1-z)}
  +\pi^2\*\L p(z)+{4\over 3}\*(1+z) \R \bigg\},
  \label{eq:Kr1}
\end{eqnarray}
with $p(z)$ defined in \Eq{eq:PolSumLO}, and where~\cite{Vo96}
\begin{equation}
 S_2(z) = \int_{{z\over 1+z}}^{{1\over 1+z}} {dw\over w} 
   \ln \biggl[{1-w\over w}\biggr]
= -2 \Li2(-z) -2 \ln z \ln (1+z) + {1\over2} \ln^2 z
                        -{\rlap{$\pi^2$}\pi\over6}.
\end{equation}
In the derivation of \Eq{eq:Kr1} we have used the integrals given 
in the appendix.

Next, we compute the two additional contributions from the
second and third terms in~\Eq{eq:ContributingTripleCollinearRegions}.
The limiting function which we need to integrate is given by \cite{CaGl98}
\begin{equation}
  \Mixed(i \soft, j\parallel k) = 2\*{(1-z_j+z_j^2)^2\over (1-z_j)\*z_j\*s_{jk}  } 
  \*{(z_j\*s_{jk}+z_j\*s_{ijk}+s_{ij})\over 
    s_{ij}\*s_{ijk}}{s_{jl}+s_{kl}\over s_{il}}, 
\end{equation}
with $l$ being the momenta of the adjacent color connected hard parton
in the antenna containing the soft gluon.
(The normalization is again different from 
\cite{CaGl98} on account of different normalization conventions for the color
matrices.) 
The term 
\def\outdent{\hskip -10mm}
\begin{eqnarray}
  && \outdent\ThetaS(1;2,3;4)\*(\ThetaW(1,2,3)-\Theta(\smin-s_{123}))
  \, \Mixed(1 \soft, 2\parallel 3)
\nn\\&&
   \outdent =\Theta(s_{34}-\smin)\Theta(s_{24}-\smin)
  \*\Theta(\smin-s_{14})\*(\ThetaW(1,2,3)-\Theta(\smin-s_{123}))
  \, \Mixed(1 \soft, 2\parallel 3)
\end{eqnarray}
in \Eq{eq:ContributingTripleCollinearRegions} yields the following contribution,
\begin{eqnarray}
  &&
  6 \gs^4 N^4 \int \dlips(1,2,3,4) |{A}^{(0)}(1,2,3,4)|^2
 \sum_{i=1}^4 \delta(x_g - x_i)\nn\\
 &&\Theta(s_{34}-\smin)\Theta(s_{24}-\smin)
  \*\Theta(\smin-s_{14})\*(\Theta_{b}(1,2,3)-\Theta(\smin-s_{123}))
  \, \Mixed(1 \soft, 2\parallel 3)\nn\\
  && =
  {1\over 4}\L \as\over2\pi\R ^2 N^2
  \int dz \int \dlips(P,4)\;
  |{A}^{(0)}(P,4)|^2
  \delta(x_g - z x_P) K_\delta(z),
\end{eqnarray}
where
\begin{eqnarray}
  && K_\delta(z)= 128\pi^4 N^2 \int \dRcoll(1,2,3)
 (\delta(z-z_2) + \delta(z-(1-z_2)))\Mixed(1 \soft,2\parallel 3)\nn\\
 &&\phantom{K_\delta(z)=}
  \times\Theta(s_{34}-\smin)\Theta(s_{24}-\smin)
  \*\Theta(\smin-s_{14})\*(\ThetaW(1,2,3)-\Theta(\smin-s_{123})).
\end{eqnarray}
The easiest way to obtain $K_\delta(z)$ is to calculate the
contributions from $\ThetaW(1,2,3)$ and $\Theta(\smin-s_{123})$ 
separately and then take the difference. The integration over 
the region given by $\Theta(\smin-s_{123})$ is similar to that
which yields~\Eq{eq:K2real}. 
For the integration over the region given by $\ThetaW(1,2,3)$ we can use the
result given in \cite{GeGl98}. 
Subtracting the two contributions the final result reads:
\begin{eqnarray}
  K_\delta(z) = &&
  - \* 2 N^2
   \*\L {4\*\pi\mu^2\over \smin} \R ^{2\e} \L {s_{P4}\over \smin}\R^\e 
  \*{1\over \Gamma(1-2\*\e)} p(z)\nn\\ 
  &&\times \*{1\over \e^3} \* \bigg( 
  {\Gamma(1-2\*\e)\over\Gamma(1-\e)^2}\*(1-z)^{-\e}
  +{\Gamma(1-2\*\e)\over\Gamma(1-\e)^2}\*(z)^{-\e}
  -2\* z^{-\e} \*(1-z)^{-\e}
  \bigg). 
\end{eqnarray}
One can also compute the integral directly, as a check, and 
we obtain the same result.
Including all other terms with the same structure ($3\soft, 1\parallel 2$ and
cyclic permutations),
we finally obtain
\begin{equation}
   \L \as^2\over2\pi\R ^2 N^2
  \int dz \int \dlips(P,4)\;
  |{A}^{(0)}(1,2)|^2
  [\delta(x_g - z x_1)+\delta(x_g - z x_2)] K_\delta(z).
  \label{eq:TheWedge}
\end{equation}
Combining the separate contributions from
\Eq{eq:OneCollinearPair}, \Eq{eq:TwoIndependentCollinearGluons},
\Eq{eq:TripleCollinear}, and 
\Eq{eq:TheWedge} our final result for the singular part
of the four parton final state is,
\begin{eqnarray}
   \L{\as\over 2\pi}\R^2 \bigg( {1\over \e} \hr1\otimes\K0
  + {1\over \e } \h0\otimes \L \Kr1
  - {1\over \e} \K0 \otimes \K0 \R 
  + {1\over \e} \cV(z)
  \bigg)
\label{eq:FourPartonRealSingular}
\end{eqnarray}
in which,
\begin{eqnarray}
  &&-2 (\cS^{(0)}(1,2) + \cC^{(0)}(1,2,1)) {\as\over 2\pi} {1\over \e}
  \K0(z)
  +\cC^{(0)}(z 1,2,(1-z)1)  {\as\over 2\pi}{1\over \e} \K0(z)
  +\L{\as^2\over2\pi}\R ^2 K_\delta(z)\nn\\
  &=&
  -\L {\as\over 2\pi}\R ^2
  \* N^2\* {1\over\Gamma(1-\e)^2}\*
  \L{4\*\pi\*\mu^2\over \smin}\R ^{2\e}  {1\over \e^3} 
  2 \*(z\*(1-z))^{-\e}\* p(z) 
  \L -2 - {11 \over 6}\*\e +({2 \over 3} \*\pi^2-{67 \over 18})\*\e^2
  + O(\e^3) \R \nn\\
  &\equiv&\L {\as\over 2\pi}\R ^2 {1\over \e^2} \cV(z).
  \label{eq:DefinitionV}
\end{eqnarray}
\subsection{Assembling the Kernel}
\label{sec:Assembly}
Upon combining the results for the singular contribution of three- and
four-parton final states, \Eq{eq:ThreePartonVirtualSingular} 
and \Eq{eq:FourPartonRealSingular} respectively,
we obtain
\begin{eqnarray}
  {1\over \e}  (\hv1 + \hr1) \otimes \K0 
  +  
  \h0 \otimes \L {1\over \e} \Kr1 
  -{1\over \e} \K0 \otimes {1\over \e} \K0  
  + {1\over \e^2} \cV
  +{1\over \e} \Kv1
  \R. 
\end{eqnarray}
Note that the equation above contains only $1/\e^2$ and $1/\e$
singularities, the $1/\e^3$ poles present in individual terms cancel
in the sum.
In addition, the singularity is independent of $\smin$ as it ought to be.
Inserting this result in \Eq{eq:MSbarCounterterm} we obtain
\begin{eqnarray}
  {d\GRs\over dx_g} &=&
  {1\over \e}  \h1 \otimes \K0
  +  
  \h0 \otimes \L {1\over \e} \Kr1 
  -{1\over \e} \K0 \otimes {1\over \e} \K0  
  + {1\over \e^2} \cV
  +{1\over \e} \Kv1
  \R\nn\\
  &+&  \h1 \otimes {1\over \e} \P0
  + \h0\otimes \L 
   {1\over 2\e^2} \Se^2 \P0\otimes \P0
  -{1\over 4\e^2} \Se^2 \beta_0 \P0
  +{1\over 2}\* {1\over \e} \Se^2 \P1\R + {\cal O}(\e^0) \nn\\
  &=&
  \h0 \otimes \L 
  {1\over \e} \Kr1 
  +{1\over \e} \Kv1
  -{1\over 2}{1\over \e} \K0 \otimes {1\over \e} \K0  
  +{1\over \e^2} \cV
  -{1\over 4\e^2} \Se^2 \beta_0 \P0
  +{1\over 2}\*{1\over \e} \Se^2 \P1 \R \nn\\
  &+& \h1 \otimes {1\over \e} (\Se\P0+\K0)\nn\\
  &+& \h0\otimes \L 
  {1\over 2\e^2} \Se^2 \P0\otimes \P0
  -{1\over 2\e^2} \K0 \otimes  \K0  
  \R 
  + {\cal O}(\e^0).
\end{eqnarray}
Thanks to the leading order result in~\Eq{eq:LeadingOrderResult}, we see that
\begin{equation}
  \hR1 = \h1 + {1\over \e}\h0 \otimes  \Se\P0
\end{equation}
is finite, so that we may write,
\begin{eqnarray}
   {d\GRs\over dx_g} &=&
  \h0 \otimes \L 
  {1\over \e} \Kr1 
  +{1\over \e} \Kv1
  -{1\over 2\e^2} \K0 \otimes  \K0  
  +{1\over \e^2} \cV
  -{1\over 4\e^2} \Se^2 \beta_0 \P0
  +{1\over 2}\* {1\over \e} \Se \P1 \R\nn\\
   &+& 
  \hR1 \otimes {1\over \e} (\Se\P0+\K0)\nn\\
  &-&
  \h0\otimes {1\over 2\e^2} \L 
   \Se^2\P0\otimes \P0
  + 2 \Se\P0\otimes \K0  
  + \K0 \otimes \K0  
  \R 
  + \cO(\e^0).
  \label{eq:PutItTogether}
\end{eqnarray}
The term involving $\hR1$ is finite as it should be. 
The last term can be written in a more suggestive form,
\begin{eqnarray}
  - \h0\otimes {1\over 2\e^2} (\Se\P0+ \K0) \otimes (\Se\P0+ \K0) .
  \label{eq:PMinusKSquare}
\end{eqnarray}
The leading-order result implies that
\begin{equation}
  \Se\P0+ \K0 
\end{equation}
is of order $\e$, and the convolution will not produce additional
singularities. This has been checked via an explicit calculation. 
We may therefore drop the term in 
\Eq{eq:PMinusKSquare}.
The requirement that the remaining, first, term on the right-hand side
of \Eq{eq:PutItTogether} be finite then allows us to extract the next-to-leading
order kernel $\P1$,
\begin{equation}
  {1\over 2}\*\Se^2 \P1 = 
  \L 
  -\Kr1 
  -\Kv1
  +{1\over 2\e} \K0 \otimes  \K0  
  -{1\over \e} \cV(z)
  +{1\over 4\e} \Se^2 \beta_0 \P0
  \R.
\end{equation}
Intuitively, the first two terms are in some sense 
the radiative corrections to the leading-order splitting amplitude, while
the following two terms,
\begin{equation}
 {1\over 2\e} \K0 \otimes  \K0    - {1\over \e}\* \cV(z)
\end{equation}
just remove the iteration of the leading-order kernel.  The last
term can be thought of as an ultraviolet renormalization.
Plugging in the explicit results for $\Kr1$, $\Kv1$, $\cV$
[\Eq{eq:Kv1}, \Eq{eq:Kv1ct}, \Eq{eq:Kr1}, \Eq{eq:DefinitionV}], 
together with 
\begin{eqnarray}
  [\K0\otimes\K0](z)   &=&
  (16\*\pi^2 \*\collinearNorm\* N)^2
  \*\bigg(\delta(1-z)\*
  \cN^2
  +{22 \over 3} \*p(z)
  +12-{44 \over 3} \*{1\over z}-12\*z+{44 \over 3} \*z^2\nn\\
  &&
  -8\*
  \ln(z)
  \*p(z)
  +4\*{(1-4\*z+3\*z^2+z^4)\over z\*(1-z)}\*\ln(z)
  +8\*\ln(1-z)\*p(z)
  \nn\\
  &+ &
  \e\*\bigg[
  -{2 \over 9} \*{(1-z)\over z}\*(67-2\*z+67\*z^2)
  +{4 \over 3} \*\pi^2\*{1+3\*z^2-4\*z^3+z^4\over z\*(1-z)}
  -16\*(1+z)\*\Li2(z)\nn\\
  &&
  +{8 \over 3} \*{(1-z)\over z}\*(11+2\*z+11\*z^2)\*\ln(1-z)
  -12\*p(z)\*\ln(1-z)^2\nn\\
  &&
  +{4 \over 3} \*{1\over z}\*(11+3\*z+12\*z^2)\*\ln(z)
  +8\*(1+z)\*\ln(z)^2\nn\\
  && +2\*(2 \*\ln(z)^2
  -{11 \over 3} \*\ln(z)
  -{11 \over 3} \*\ln(1-z)
  -{2 \over 3} \*\pi^2
  +{67 \over 9} )\*p(z) 
  \bigg]
  \bigg)+O(\e^2),
  \label{eq:KxK}
\end{eqnarray}
we finally obtain,
\begin{eqnarray}
  \P1(z)
  &=&
  N^2\* \bigg({27 \over 2} \*(1-z) 
  +{67 \over 9} \*(z^2-{1\over z}) 
  + ({11 \over 3}  - {25 \over 3} \*z-{44 \over 3} \*{1\over z})\*\ln(z) 
  - 4\*(1+z)\*\ln(z)^2\nn\\
  &+&(4\*\ln(z)\*\ln(1-z) 
  -3 \*\ln(z)^2 + {22 \over 3} \*\ln(z)
  -{1 \over 3} \*\pi^2 + {67 \over 9} ) \* p(z)
  + 2\* p(-z)\*S_2(z)\bigg)
\end{eqnarray}
in agreement with known results for the time-like 
kernel~\cite{CuFuPe80,FlRoSa77,FlRoSa79}. 

\section{Conclusion}
\label{sec:Conclusion}

Intuitively, the Altarelli--Parisi kernel summarizes that part
of collinearly unresolved radiation from a short-distance process which
must be absorbed into the scaling evolution of descriptions of initial-
or final-state hadrons.  The computation described above makes this
direct connection precise, and shows how to use it in order to compute
the kernel.  The approach described in the present paper effectively
breaks down the NLO computation into smaller and simpler parts, whose
intermediate terms have an independent meaning and are subject to 
consistency checks on their own.  

As described in section~\ref{sec:Assembly}, the approach effectively
computes the the next-to-leading order kernel as a radiative correction
to the leading-order kernel (after cancellation of soft singularities),
less an iteration of the leading-order kernel.  As usual, the radiative
corrections arise from a virtual correction and a real-emission correction.
Both are computed from higher-order splitting amplitudes, quantities
which govern the collinear behavior of gauge-theory amplitudes.  These
intermediate quantities are useful elsewhere in their own right,
and can be checked independently.  For example, the virtual corrections
are basically given by the one-loop splitting amplitude, which should
satisfy certain supersymmetry identities~\cite{KoUw99a}.  The real-emission
correction arises from integrating the splitting amplitude describing
the behavior of tree-level amplitudes in a gauge theory as three color-connected
partons become collinear. 

The complete computation of the NNLO corrections to the Altarelli--Parisi
kernel remains an important goal for particle theorists.   These
corrections, and parton distribution functions relying on them, are
needed for programs evaluating jet production to NNLO at hadron
colliders. In the approach described
in this paper, the required ingredients would be the two-loop $1\rightarrow 2$
splitting amplitudes; the one-loop $1\rightarrow 3$ splitting amplitudes;
and the $1\rightarrow 4$ splitting amplitudes.  
The second ingredient has been calculated by Catani, de~Florian and
Rodrigo \cite{CaFl03} 
and the third by Del Duca, Frizzo and Maltoni \cite{DelDuca:1999ha}.

{\bf Acknowledgments:} 
One of us (P.U.) would like to thank Arnd Brandenburg, Aude
Gehrmann-de~Ridder, Gudrun
Heinrich, Sven Moch, and  Stefan Weinzierl for useful
discussions.  We would also like to thank Zvi Bern, Sven
  Moch, and Stefan Weinzierl for a careful reading of the manuscript.

\appendix
\renewcommand\theequation{\ifnum \value{section}>0
 \Alph{section}.\arabic{equation}%
\else
\arabic{equation}%
\fi}

\section{Double unresolved phase space integration}
\label{sec:PhaseSpaceIntegrals}
The calculation of contributions from the four-parton  final state 
in particular \Eq{eq:K2real} leads
to the consideration of integrals of the following form 
\begin{equation}
  C_{n,m} = s_{ijk}^{-(2-n-m)} 
  \int\limits_{-\Delta>0} \!\!\!
  ds_{ij}ds_{jk}\,
  {1\over s^n_{ij}s^{m}_{jk}} (-\Delta)^{-{1\over 2}-\e} 
\end{equation}
with 
\def\tzi{z_i}
\def\tzj{z_j}
\def\tzk{z_k}
\begin{equation}
  \Delta = (s_{ik}\*(1-\tzi-\tzk)- \tzi\*s_{jk}-s_{ij}\*\tzk)^2 
  -4\*\tzk\*\tzi\*s_{jk}\*s_{ij}.
\end{equation}
The integration can be done in $d=4-2\e$ dimensions yielding
\begin{eqnarray}
  C_{1,1} 
  &=&
  -2 \* \pi \* {1\over \e}\*s_{ijk}^{-1-2\*\e}
  \*\tzi^{-\e}
  \*\tzk^{-\e}
  \*\tzj^{-1-2\*\e}
  \*(1-\tzi)^\e
  \* (1-\tzk)^\e 
  \*\F21(-\e,-\e,1-\e,{\tzi\over 1-\tzi}\*{\tzk\over 1-\tzk}),\nn\\
  C_{1,0} 
  &=&
  - \pi \* {1\over \e}\*s_{ijk}^{-1-2\e}
  \*\tzi^{-\e}
  \*\tzk^{-\e} 
  \*\tzj^{-\e}
  \*{1\over 1-\tzk}
  ,\nn\\
  C_{1,-1} 
  &=&
  - \pi \* {1\over \e}\*s_{ijk}^{-1-2\*\e}\*{1\over 1-2\*\e}
  \*\tzi^{-\e}
  \*\tzk^{-\e}
  \*\tzj^{1-\e}
  \*{1\over(1-\tzk)^2} 
  \*(1-\e-\e\*{\tzi\*\tzk\over \tzj}),\nn\\
  C_{0,1} 
  &=&
  - \pi \* {1\over \e}\*s_{ijk}^{-1-2\*\e}
  \*\tzi^{-\e}
  \* \tzj^{-\e}
  \*\tzk^{-\e}\* (1-\tzi)^{-1}
  ,\nn\\
  C_{0,0} 
  &=&
   \pi \* {1\over \e}\*s_{ijk}^{-1-2\*\e}
  \*{\e\over 1-2\*\e}
  \*\tzi^{-\e}
  \*\tzj^{-\e}
  \*\tzk^{-\e},
\end{eqnarray}
and
\begin{eqnarray}
  \lefteqn{
    z_b^2 C_{2,0}
    -2z_b(1-z_c) C_{2,-1}
    +(1-z_c)^2 C_{2,-2}
    }\nn\\
  &=&
   \pi \* {1\over \e}\*
  {1 \over 1-2\*\e }
  \*s_{abc}^{-1-2\*\e}
  \*z^{-\e}_a\*z^{-\e}_b\*
  \*z^{-\e}_c\*(1-z_a)^{2}
  \L\e-{z_b\*(2+4\*\e)\over z_b+z_a\*z_c}
  +{z_b^2\*(2+4\*\e)\over (z_b+z_a\*z_c)^2}
  \R \label{eq:C2mIntegrals}.
\end{eqnarray}
Note that the $C_{2,m}$ integrals can not appear in arbitrary
combinations in a gauge theory amplitude. The individual integrals
appearing in \Eq{eq:C2mIntegrals} are not regularized in dimensional
regularization, in contrast the specific combination is regularized.
The combination given above is exactly the combination appearing in 
\Eq{eq:K2real}.

\def\twoloopfourpoint{%
C.\ Anastasiou, E.\ W.\ N.\ Glover, C.\ Oleari, and M.\ E.\ Tejeda-Yeomans, 
Nucl.\ Phys.\ B601 (2001) 318,
\href{http://www.arXiv.org/abs/hep-ph/0010212}{{\tt hep-ph/0010212}};
C.\ Anastasiou, E.\ W.\ N.\ Glover, C.\ Oleari, and M.\ E.\ Tejeda-Yeomans,
Nucl.\ Phys.\ B601 (2001) 341, \href{http://www.arXiv.org/abs/hep-ph/0011094}{{\tt hep-ph/0011094}};
C.\ Anastasiou, E.\ W.\ N.\ Glover, C.\ Oleari, and M.\ E.\ Tejeda-Yeomans,
Nucl.\ Phys.\ B605 (2001) 486, \href{http://www.arXiv.org/abs/hep-ph/0101304}{{\tt hep-ph/0101304}};
E.\ W.\ N.\ Glover, C.\ Oleari, and M.\ E.\ Tejeda-Yeomans,
Nucl.\ Phys.\ B605 (2001) 467, \href{http://www.arXiv.org/abs/hep-ph/0102201}{{\tt hep-ph/0102201}};
L.\ W.\ Garland, T.\ Gehrmann, E.\ W.\ N.\ Glover, A.\ Koukoutsakis, and E.\ Remiddi,
Nucl.\ Phys.\ B627 (2002) 107, \href{http://www.arXiv.org/abs/hep-ph/0112081}{{\tt hep-ph/0112081}};
L.\ W.\ Garland, T.\ Gehrmann, E.\ W.\ N.\ Glover, A.\ Koukoutsakis, and E.\ Remiddi,
Nucl.\ Phys.\ B642 (2002) 227,  \href{http://www.arXiv.org/abs/hep-ph/0206067}{{\tt hep-ph/0206067}};
S.\ Moch, P.\ Uwer, and S.\ Weinzierl,
Phys.\ Rev.\ D66 (2002) 114001, 
\href{http://www.arXiv.org/abs/hep-ph/0207043}{{\tt hep-ph/0207043}};
Z.\ Bern, A.\ De Freitas, and L.\ Dixon,
\href{http://www.arXiv.org/abs/hep-ph/0211344}{{\tt hep-ph/0211344}} 
}

\def\allplus{%
Z.\ Bern, L.\ J.\ Dixon and D.\ A.\ Kosower,
\href{http://www.arXiv.org/abs/hep-th/9311026}{{\tt hep-th/9311026}};
Z.\ Bern, G.\ Chalmers, L.\ J.\ Dixon and D.\ A.\ Kosower,
Phys.\ Rev.\ Lett.\ 72 (1994) 2134
\href{http://www.arXiv.org/abs/hep-ph/9312333}{{\tt hep-ph/9312333}};
Z.~Bern, A.~De Freitas and L.~J.~Dixon,
JHEP {0109} (2001) 037 
\href{http://www.arXiv.org/abs/hep-ph/0109078}{{\tt hep-ph/0109078}};
Z.~Bern, A.~De Freitas and L.~Dixon,
JHEP {0306} (2003) 028 
\href{http://www.arXiv.org/abs/hep-ph/0304168}{{\tt hep-ph/0304168}}
}
\newpage

\newcommand{\zp}{Z. Phys. }\def\as{\alpha_s }\newcommand{\prd}{Phys. Rev.
  }\newcommand{\pr}{Phys. Rev. }\newcommand{\prl}{Phys. Rev. Lett.
  }\newcommand{\npb}{Nucl. Phys. }\newcommand{\psnp}{Nucl. Phys. B (Proc.
  Suppl.) }\newcommand{\pl}{Phys. Lett. }\newcommand{\ap}{Ann. Phys.
  }\newcommand{\cmp}{Commun. Math. Phys. }\newcommand{\prep}{Phys. Rep.
  }\newcommand{\jmp}{J. Math. Phys. }\newcommand{\rmp}{Rev. Mod. Phys. }

\end{document}